\documentclass[10pt,a4paper]{article}

\usepackage[a4paper,top=20mm,bottom=20mm,left=18mm,right=18mm,
footskip=9mm]{geometry}
\usepackage[T1]{fontenc}
\usepackage[utf8]{inputenc}
\usepackage{amsmath,amsthm}
\IfFileExists{newtxtext.sty}{\usepackage{newtxtext,newtxmath}}{\usepackage{mathptmx}\usepackage{amssymb}}
\usepackage{microtype}
\usepackage[numbers,sort&compress]{natbib}
\usepackage{booktabs,tabularx,array,multirow}
\usepackage{titlesec}
\usepackage{caption}
\usepackage{subcaption}
\usepackage{graphicx}
\usepackage{float}
\usepackage{placeins}
\usepackage{xcolor}
\usepackage{pifont}
\usepackage{algorithm}
\usepackage{algpseudocode}
\usepackage{tikz}
\usetikzlibrary{positioning,arrows.meta,shapes.geometric,fit,backgrounds,calc,decorations.pathreplacing}
\usepackage{pgfplots}
\pgfplotsset{compat=1.17}
\usepackage[hidelinks]{hyperref}

\titlespacing*{\section}{0pt}{8pt}{3pt}
\titleformat{\section}{\normalfont\bfseries\large}{\thesection.}{0.5em}{}
\titlespacing*{\subsection}{0pt}{6pt}{2pt}
\titleformat{\subsection}{\normalfont\bfseries\normalsize}{\thesubsection}{0.5em}{}
\titlespacing*{\subsubsection}{0pt}{4pt}{1.5pt}
\titleformat{\subsubsection}{\normalfont\bfseries\itshape\normalsize}{\thesubsubsection}{0.5em}{}
\titleformat{\paragraph}[runin]{\normalfont\bfseries\normalsize}{}{0pt}{}
\newcolumntype{L}[1]{>{\raggedright\arraybackslash}p{#1}}

\newcommand{\dataset}{\textit{SnaanGhar7}}
\newcommand{\meanstd}[2]{#1\,$\pm$\,#2}

\let\Oldincludegraphics\includegraphics
\renewcommand{\includegraphics}[2][]{%
	\IfFileExists{#2}{\Oldincludegraphics[#1]{#2}}{%
		\fbox{\parbox[c][0.22\textheight][c]{0.82\linewidth}{\centering Figure placeholder:\\ \texttt{\detokenize{#2}}}}}%
}

\newcommand{\sep}{\unskip;\ }

\begin{document}

	\begin{center}
		{\Large\bfseries SincDPNet: Interpretable Raw-Waveform Bathroom Activity Recognition for Assistive Living}\\[10pt]

		Debolina Chowdhury$^{1}$, Suman Samui$^{2,*}$, Sujoy Saha$^{1}$\\[4pt]
		{\small\itshape $^{1}$Department of Computer Science and Engineering, National Institute of Technology Durgapur,\\
			Durgapur 713209, West Bengal, India}\\[2pt]
		{\small\itshape $^{2}$Department of Electronics and Communication Engineering, National Institute of Technology Durgapur,\\
			Durgapur 713209, West Bengal, India}\\[4pt]
		{\small Emails: \texttt{debolinarimuchowdhury@gmail.com} (D.~Chowdhury), \texttt{ssamui.ece@nitdgp.ac.in} (S.~Samui),\\
			\texttt{ssaha.cse@gnitdgp.ac.in} (S.~Saha)}\\[2pt]
		{\small $^{*}$Corresponding author}
	\end{center}
	\vspace{2pt}

\section*{Abstract}
	Bathroom acoustic-event recognition can support ambient assisted living in
	settings where continuous video monitoring is undesirable. However, practical
	deployment requires models that are compact, interpretable, and robust to
	changes in the recording environment. This work introduces \dataset{}, a
	seven-class bathroom acoustic-event dataset containing 21{,}387 annotated
	clips recorded across five environments, and proposes SincDPNet, a compact
	raw-waveform classifier with a learnable sinc filter bank followed by a
	depthwise-separable convolutional body. Each sinc filter is controlled by two
	frequency parameters, allowing the learned passbands to be inspected directly
	in hertz while keeping the front end small. To reduce room-specific leakage,
	recording sessions and environments are separated before overlapping windows
	are assigned to the training, validation, and test partitions. We further use
	multi-objective Bayesian optimization as a design tool to examine the
	validation performance--model-size trade-off across 24 configurations. The
	selected designs span different operating points: the best-performing model
	achieves 80.2\% accuracy and 0.760 macro-F1 with 14{,}040 parameters, while
	the compact $N_f=25$ configuration uses only 2{,}848 parameters and achieves
	75.7\% accuracy, 0.661 macro-F1, and 0.716 MCC on the held-out environment.
	Analysis of the learned filters and confusion patterns shows that spectral
	overlap contributes to confusion among water-related events, while the
	\textit{Door}/\textit{Walker/Crutch} errors also reflect similarities in
	their transient temporal structure.

\vspace{4pt}
\noindent\textbf{Keywords:} Acoustic scene classification \sep Bathroom acoustic event recognition \sep Raw-waveform learning \sep
    SincNet \sep Interpretable deep learning \sep Bayesian optimization \sep Ambient assisted living \sep TinyML \sep Edge AI

\section{Introduction}
\label{sec:intro}

Acoustic event classification (AEC) provides a means of recognizing human
activities and environmental context without requiring wearable sensors or
visual monitoring~\citep{hershey2017cnn,gemmeke2017audioset}. Acoustic
sensing has therefore been explored as a less visually intrusive option for
indoor activity monitoring, particularly in privacy-sensitive
environments~\citep{marmaroli2023indoor}. Bathrooms are an important example:
they are private spaces where continuous camera-based monitoring is generally
unacceptable~\citep{goetze2022audio,tong2024acceptability}, yet information
about activities such as flushing, showering, tap use, door movement, and
walker or crutch use can be relevant to Ambient Assisted Living (AAL).

This setting is particularly relevant for older adults living independently,
for whom bathroom activities may be associated with slips, falls, or delayed
assistance. A sequence of acoustic events can provide a simple account of
bathroom activity without continuous video capture. For example, combinations
of door, mobility-aid, tap, shower, and flush events can indicate the progress
of a routine and, in a future assistive system, provide evidence of prolonged
inactivity or unusual activity patterns.

Several practical difficulties make this problem less straightforward than
general environmental sound classification. Suitable public datasets are
limited. Widely used corpora such as ESC-50~\citep{piczak2015esc},
UrbanSound8K~\citep{salamon2014dataset},
AudioSet~\citep{gemmeke2017audioset},
FSD50K~\citep{fonseca2022fsd50k}, and the TAU acoustic scene
datasets~\citep{mesaros2018multi} contain few or no recordings targeted at
bathroom activities, while many studies in this area rely on private data.
The second challenge relates to model interpretability. Standard
spectrogram-based approaches can obtain high accuracy on recognition, yet
the learned representation does not provide any information regarding the
acoustic characteristics that can be used to differentiate, say, flushing a toilet
and turning on a faucet. Interpretability of such mistakes becomes very
relevant in assistive monitoring, where one should be able to explain the
mistake in terms of the actual sound. Resource-efficient implementation,
however, poses another constraint: accuracy should be traded off against
model complexity.

Accordingly, this work makes the following contributions:

\begin{enumerate}
	
	\item \textbf{The \dataset{} dataset.}
	We consolidate and publicly release a real-world, multi-site
	assistive-living bathroom acoustic corpus containing seven
	activity-related classes, including a heterogeneous non-target
	\emph{unknown} class. The dataset contains 21{,}387 annotated clips,
	standardized data splits, and baseline implementations
	(Section~\ref{sec:dataset}). Its scope is compared with the closest
	existing bathroom and environmental acoustic resources in
	Section~\ref{sec:related}.
	
	\item \textbf{The SincDPNet architecture.}
	We develop a compact raw-waveform model that combines a learnable sinc
	band-pass front end with a depthwise-separable convolutional back end
	(Section~\ref{sec:arch}). Each sinc filter is represented by two
	frequency parameters, giving a filter-bank cost of only $2N_f$
	trainable parameters while retaining a direct interpretation in hertz.
	SincDPNet differs from SincNet \cite{ravanelli2018speaker}\cite{Mittermaier2020} in the architecture following the sinc
	layer: the parameter-intensive convolutional and classification stages \cite{Mittermaier2020}
	are replaced by a compact separable back end. The resulting model retains
	the frequency interpretability of the sinc representation while reducing
	the network to a few thousand trainable parameters.
	
	\item \textbf{Acoustic analysis and reproducible design exploration.}
	We examine acoustic overlap between classes before model training and
	relate this structure to the errors observed after training. The analysis
	shows that pairwise acoustic overlap is informative about several dominant
	confusion patterns, although a single class-level difficulty score does
	not predict the complete error ranking. We further report the
	multi-objective design study using all 24 evaluated configurations
	(Section~\ref{sec:mobo}), allowing the resulting accuracy--size trade-off
	to be inspected directly.
	
\end{enumerate}

The optimization procedure itself is used as a design tool; no new Bayesian
optimization method is introduced. Similarly, \dataset{} is presented as a
new resource for the particular combination of bathroom activities studied
here, rather than as the first bathroom acoustic dataset. The contribution of
the study lies in the combination of an interpretable compact raw-waveform
model, an acoustic analysis that can be related to its learned filter bank and
error structure, and a dataset that permits these relationships to be studied
under environment-disjoint evaluation.

The remainder of the paper is organized as follows.
Section~\ref{sec:related} reviews related work on bathroom and environmental
acoustic datasets, spectral and raw-waveform front ends, compact edge-audio
models, and multi-objective optimization. Section~\ref{sec:dataset}
introduces \dataset{}, including its recording protocol, annotation, and
statistics. Section~\ref{sec:characterization} examines the signal-level
structure and inter-class acoustic similarity. Section~\ref{sec:arch}
describes SincDPNet and its relationship to SincNet, and
Section~\ref{sec:mobo} presents the multi-objective design study.
Section~\ref{sec:results} reports the benchmark, ablation, interpretability,
and error analyses, followed by the deployment evaluation in
Section~\ref{sec:deployment}. The findings are discussed in
Section~\ref{sec:discussion}, and Section~\ref{sec:conclusion} concludes the
paper.
\section{Related Work}
\label{sec:related}

Literature related to this work is divided into the following five categories, which are interrelated:
dataset of bathroom and environmental acoustics, spectral and waveform
signal front-end, compressed models for edge audio, multi-objective
design, and evaluation in case of domain shift. All these categories
represent real constraints associated with monitoring bathroom acoustics,
namely limited data, similar acoustic classes, computational constraint,
and evaluation outside of training domains.

\subsection{Environmental and Bathroom Acoustic Datasets}

Progress in acoustic event classification has been supported by large
general-purpose corpora such as ESC-50~\citep{piczak2015esc},
UrbanSound8K~\citep{salamon2014dataset}, and
AudioSet~\citep{gemmeke2017audioset}. These datasets contain a variety of sounds from urban, domestic, and everyday settings, however, bathroom-related sound events are missing or not well covered.

Bathroom acoustic sensing has been studied for considerably longer than the
size of the literature might suggest. Chen et al.~\cite{chen2005bathroom}
showed that flushing, showering, and tap use could be distinguished from
audio alone. More recently, Hyun~\cite{hyun2024sound} considered an
edge-oriented system for three water-usage activities, using YAMNet for
water-event detection followed by a fine-tuned classifier on a Raspberry Pi.
\"Ozt\"urk et al.~\cite{ozturk2025novel} released a public restroom corpus of
approximately 460 minutes recorded in five bathrooms and reported 97.8\%
accuracy over eleven fine-grained events using a RegNetY backbone with
three-channel mel spectrograms.

These studies establish the feasibility of bathroom acoustic monitoring, but
their event coverage is concentrated mainly on water-related activities.
Sounds relevant to assistive monitoring, such as door movement and
walker/crutch use, are less commonly represented, and an explicit
heterogeneous non-target class is generally absent. In addition, most
published systems rely on comparatively large spectral models, leaving the
performance of very small raw-waveform models less well characterized.
Table~\ref{tab:dataset_comparison} summarizes the position of \dataset{}
relative to representative environmental and bathroom acoustic resources.

\begin{table}[H]
	\centering
	\caption{Representative datasets for environmental sound recognition and
		bathroom monitoring. \dataset{} adds mobility-aid events, an explicit
		non-target class, and raw-waveform edge baselines.}
	\label{tab:dataset_comparison}
	\renewcommand{\arraystretch}{1.15}
	\footnotesize
	\resizebox{\textwidth}{!}{%
		\begin{tabular}{lccccc}
			\toprule
			\textbf{Dataset} & \textbf{Domain} & \textbf{Bathroom} &
			\textbf{Public} & \textbf{Non-target class} &
			\textbf{Raw-waveform baselines} \\
			\midrule
			ESC-50~\citep{piczak2015esc}
			& Environmental & \ding{55} & \checkmark & \ding{55} & \ding{55} \\
			
			UrbanSound8K~\citep{salamon2014dataset}
			& Urban & \ding{55} & \checkmark & \ding{55} & \ding{55} \\
			
			AudioSet~\citep{gemmeke2017audioset}
			& General & Partial & \checkmark & \ding{55} & \ding{55} \\
			
			Chen et al.~\cite{chen2005bathroom}
			& Bathroom & \checkmark & \ding{55} & \ding{55} & \ding{55} \\
			
			Hyun~\cite{hyun2024sound}
			& Water/bathroom & \checkmark & \ding{55} & \ding{55} & \ding{55} \\
			
			\"Ozt\"urk et al.~\cite{ozturk2025novel}
			& Restroom & \checkmark & \checkmark & \ding{55} & \ding{55} \\
			
			\dataset{} (ours)
			& Bathroom events & \checkmark & \checkmark & \checkmark & \checkmark \\
			\bottomrule
	\end{tabular}}
\end{table}

\subsection{Spectral and Raw-Waveform Front Ends}

Most acoustic classifiers use a fixed spectral representation before
classification. Mel-frequency cepstral coefficients and mel spectrograms
remain common choices~\citep{barchiesi2015acoustic}, typically followed by
convolutional or recurrent models~\citep{cakir2017convolutional,
	salamon2017deep}. These representations are effective, but their frequency
scale and analysis filters are specified before training and therefore do not
adapt directly to the target domain. Earlier work by Chu et
al.~\cite{chu2009environmental} also noted that environmental sounds often
have broadband, noise-like spectra and explored physically interpretable
time--frequency representations based on matching pursuit.

Raw-waveform models remove the need for a fixed spectral transform by
learning the first-stage representation directly from the signal. A fully
unconstrained convolution, however, requires one parameter per filter tap and
offers little direct physical interpretation. Parametric front ends provide
an alternative. SincNet~\citep{ravanelli2018speaker} constrains each
first-layer filter to a band-pass response described by two cutoff
frequencies. Related approaches learn sub-band boundaries directly from
data~\citep{kim2021multiband} or use parameterized gammatone
filters~\citep{park2020cnn}. Park and Yoo~\cite{park2020cnn}, for example,
reported improved generalization from a constrained filter shape compared
with a substantially larger unconstrained convolution. Bittner et
al.~\cite{bittner2025efficient} pursue interpretability through a different
route, using diagonal state-space models for compact raw-audio
classification.

These studies show that a properly designed front-end system could decrease the number of
free parameters while maintaining an immediate dependence on the signal.
This feature is especially important in the current problem since the learned frequency bands
could be observed in hertz and compared to the spectrum features of bathroom events.
The current research aims at analyzing if this structural approach proves effective when the
entire classifier is represented by just several thousand of parameters.

\subsection{Compact Models for Resource-Constrained Audio}

Compact audio models commonly rely on depthwise-separable convolutions,
quantization, pruning, knowledge distillation, and careful control of the
receptive field. DS-CNN~\citep{zhang2017hello} demonstrated the effectiveness
of depthwise-separable convolutions for microcontroller keyword spotting, and
the DPNet family~\citep{chowdhury2026dpnet} applies a related strategy to
bathroom sounds. Lightweight inception-style models have also been developed
for edge acoustic scene classification~\citep{pham2023lightweight}. Other
approaches reduce deployment cost through pruning and
quantization~\citep{mohaimenuzzaman2023environmental}, teacher--student
distillation~\citep{cerutti2019neural,kong2020panns}, or receptive-field
design~\citep{koutini2021receptive}.

The DCASE low-complexity acoustic-scene tasks provide a useful reference for
such constraints. The 2021 edition~\citep{martinmorato2021low} limited model
size to $128$\,kB, while the 2022 edition~\citep{martinmorato2022low}
specified a budget of $128$\,k INT8 parameters and $30$ million
multiply--accumulate operations. Similar resource constraints have continued
in later editions~\citep{schmid2025low}, with successful systems making
extensive use of separable convolutions, quantization, and receptive-field
tuning~\citep{singh2022low}.

Much of this literature starts with a comparatively large architecture and
then reduces its cost. The present study instead begins with a small
parameter budget and uses a structured sinc front end together with a
depthwise-separable body. This allows model size and interpretability to be
considered jointly during design. Related work on acoustic monitoring also
shows the importance of maintaining robustness across recording
environments, including through domain adaptation~\citep{molina2024noise}.

\subsection{Multi-Objective Optimization for Edge Models}

The task of picking a particular edge model involves trade-offs between predictive accuracy and model complexity/inference cost. This is because of the dependencies between the criteria, and the natural way to formulate the design problem is through a Pareto front. Multi-objective Bayesian optimization provides a convenient tool for searching such design spaces efficiently, when evaluation of the models is expensive. In the present study, we use expected hypervolume improvement proposed by Daulton et al. ~\cite{daulton2020ehvi}. The optimization procedure is not introduced as a methodological novelty, but is used just to find operating points on a fixed budget.

\subsection{Evaluation Practice and the Gap We Address}

Model selection is only meaningful if the evaluation split reflects the
conditions expected at deployment. This issue is especially important for
small bathroom datasets, where recordings are obtained from only a few
rooms. Reverberation, microphone position, plumbing fixtures, and other
room-specific characteristics can remain consistent across many clips from
the same environment. A random clip-level split can therefore place closely
related recordings from the same room in both training and test sets,
allowing room characteristics to contribute to the reported performance.

Similar concerns are addressed explicitly in the DCASE low-complexity tasks,
where device and city information are incorporated into the evaluation
protocol~\citep{martinmorato2021low,schmid2025low}. Koutini et
al.~\cite{koutini2021receptive} likewise emphasize the role of evaluation
design in assessing generalization.

For this reason, the experiments in this study use environment-disjoint
splits, so that the environment held out for testing is not present during
training. The resulting scores are lower than those obtained with random
clip-level partitioning, but they provide a more realistic estimate of
performance in an unseen recording environment. The signal-level analysis is
used alongside this protocol to identify acoustically similar class pairs
before training and to examine whether those similarities reappear in the
model's confusion patterns. This combination of environment-disjoint
evaluation and signal-based error analysis forms the basis for the
experimental study that follows.

\section{The \dataset{} Dataset}
\label{sec:dataset}

Building on the data gap identified in Section~\ref{sec:related}, we first
describe the corpus used throughout the study.

\dataset{} is a real-world acoustic dataset developed for assistive-living
monitoring in bathrooms, with particular relevance to older adults living
independently. Bathrooms are private spaces in which several routine
activities can be monitored acoustically without continuous visual
observation. The dataset therefore focuses on activity-related sounds that
are relevant to this setting.

\begin{figure}[!htbp]
	\centering
	\includegraphics[width=0.6\linewidth]{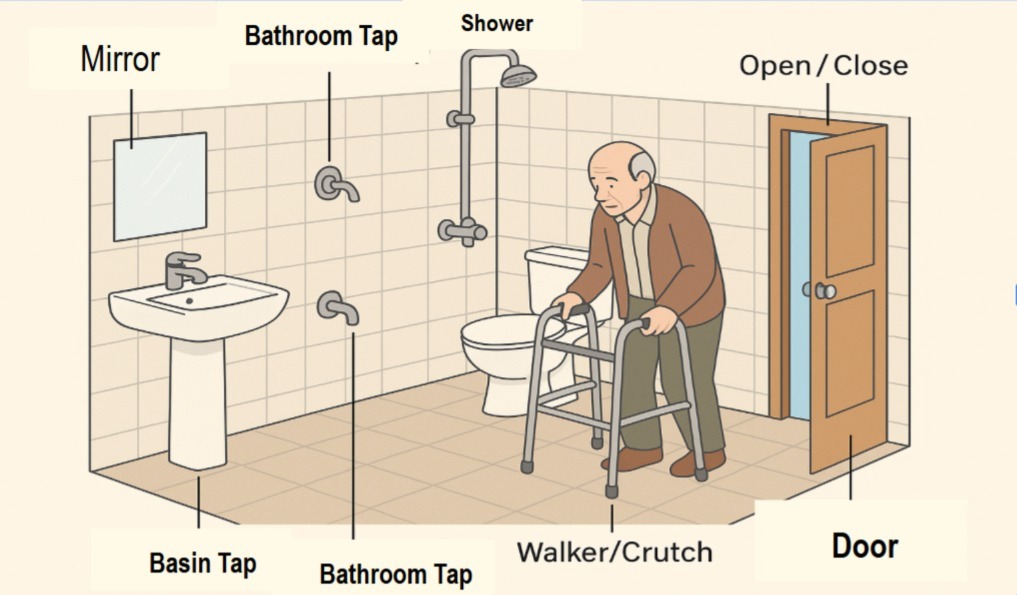}
	\caption{Assistive-living use scenario of \dataset{}. The selected sound
		classes represent bathroom activities relevant to an older adult living
		independently. Door and walker/crutch sounds provide entry, exit, and
		mobility cues, while tap, shower, and flush-related sounds correspond to
		common hygiene activities.}
	\label{fig:assistive_scene}
\end{figure}

The seven classes are \textit{Flush}, \textit{Shower},
\textit{Bathroom Tap}, \textit{Basin Tap}, \textit{Door},
\textit{Walker/Crutch}, and an \textit{Unknown Class}. These classes were
selected from an assistive-living perspective. Water-related sounds represent
common hygiene activities, \textit{Door} provides a cue for entry or exit,
and \textit{Walker/Crutch} represents mobility-aid use. When considered over
time, these events could provide a simple activity timeline and support
future systems for identifying prolonged or unusual bathroom activity.
Figure~\ref{fig:assistive_scene} illustrates the intended use scenario.

Approximately $55$ minutes of manually annotated recordings were collected
per class across five real-world bathroom environments, giving roughly
$385$ minutes of original audio. Table~\ref{tab:summary} summarizes the
dataset, while Table~\ref{tab:classes} lists the classes and their principal
acoustic characteristics.

\begin{table}[H]
	\centering
	\caption{Summary of the \dataset{} dataset and its embedded recording platform.}
	\label{tab:summary}
	\renewcommand{\arraystretch}{1.1}
	\begin{tabular}{p{0.44\linewidth}p{0.46\linewidth}}
		\toprule
		\textbf{Property} & \textbf{Value} \\
		\midrule
		Classes                 & 7 \\
		Recording environments  & 5 (residential, hostel, institutional) \\
		Contributors            & 3 \\
		Original duration       & $\approx$385 min \\
		Total segmented windows & 21{,}387 \\
		Clip duration           & 1.51125 s (30{,}225 samples) \\
		Class imbalance ratio   & $1.2\times$ \\
		\midrule
		Recording platform      & Raspberry Pi Zero~2~W \\
		Audio interface         & Waveshare WM8960 Audio HAT \\
		Sampling rate / depth   & 20 kHz / 32-bit I2S, mono \\
		\bottomrule
	\end{tabular}
\end{table}

\begin{table}[H]
	\centering
	\caption{Class definitions and dominant acoustic characteristics.}
	\label{tab:classes}
	\renewcommand{\arraystretch}{1.15}
	\footnotesize
	\setlength{\tabcolsep}{3pt}
	\begin{tabular}{p{0.16\linewidth}p{0.29\linewidth}p{0.34\linewidth}}
		\toprule
		\textbf{Class} & \textbf{Description} & \textbf{Acoustic characteristics} \\
		\midrule
		Flush         & Toilet flushing            & Broadband water flow with transient onset \\
		Shower        & Shower water flow          & Continuous broadband noise \\
		Bathroom Tap  & Bathroom tap running       & Continuous water flow, varying intensity \\
		Basin Tap     & Basin tap running          & Localized continuous water flow \\
		Door          & Opening/closing/knocking   & Short impulsive sounds \\
		Walker/Crutch & Mobility-aid movement      & Repeated transient impacts \\
		Unknown Class & Background/non-target      & Diverse environmental sounds \\
		\bottomrule
	\end{tabular}
\end{table}

\subsection{Recording Setup}

\begin{figure}[!htbp]
	\centering
	\includegraphics[width=0.6\linewidth]{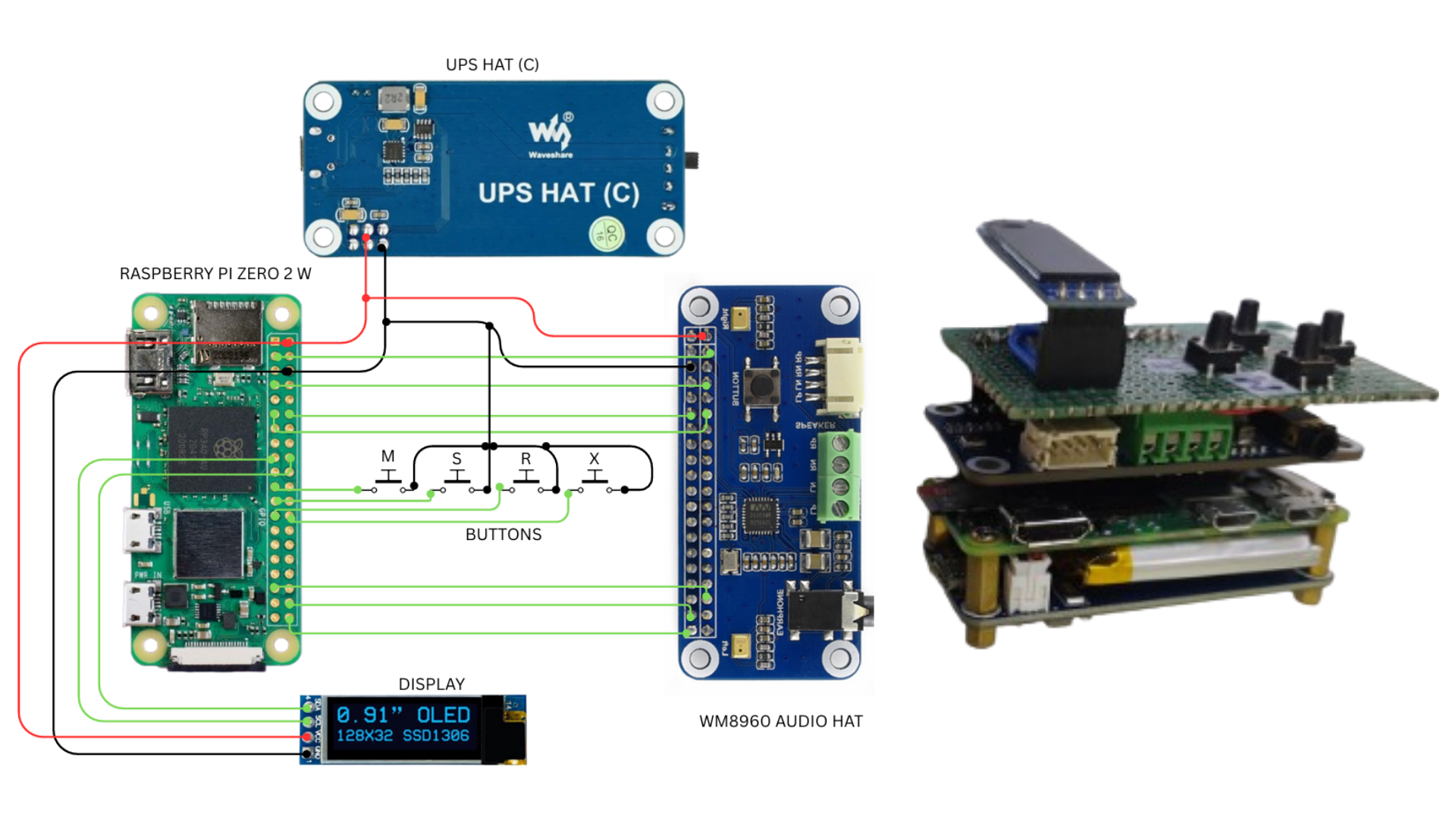}
	\caption{Embedded recording platform used for data acquisition.}
	\label{fig:prototype}
\end{figure}

Recordings were collected from five residential, hostel, and institutional
bathrooms (Table~\ref{tab:environments})\footnote{\dataset{} supports only
	activity-level acoustic classification. All recordings were collected with
	informed consent and property-owner permission. A small number of
	\textit{Unknown}-class recordings contain incidental speech captured under
	realistic conditions; these are treated solely as non-target events, and no
	personally identifiable information is included in the release.}
by three contributors using a custom embedded platform based on a Raspberry
Pi~Zero~2~W and a Waveshare WM8960 Audio HAT
(Fig.~\ref{fig:prototype}; platform details are given in
Table~\ref{tab:summary}). The five environments introduce variation in room
acoustics, reverberation, plumbing characteristics, and background noise.
This variation is used explicitly in the cross-environment evaluation
described in Section~\ref{sec:setup}.

\begin{table}[H]
	\centering
	\caption{Recording environments.}
	\label{tab:environments}
	\renewcommand{\arraystretch}{1.1}
	\begin{tabular}{p{0.10\linewidth}p{0.50\linewidth}p{0.22\linewidth}}
		\toprule
		\textbf{ID} & \textbf{Location} & \textbf{Type} \\
		\midrule
		E01 & NIT Durgapur              & Institutional \\
		E02 & Residential Home, Howrah  & Residential \\
		E03 & Hall-6, NIT Durgapur      & Hostel \\
		E04 & DS-1B, NIT Durgapur       & Residential \\
		E05 & Hall-9, NIT Durgapur      & Hostel \\
		\bottomrule
	\end{tabular}
\end{table}

\subsection{Annotation and Preprocessing}
\label{sec:preprocessing}

The continuous recordings were manually annotated at the event level and
then divided into fixed windows of $30{,}225$ samples ($1.51125$~s at
$20$~kHz) with a hop of $10{,}225$ samples, giving approximately $66\%$
overlap within each source recording. This procedure produced $21{,}387$
segmented windows. Each window is peak-normalized and centre-cropped or
zero-padded only when required to preserve the fixed input length.

\begin{figure}[!htbp]
	\centering
	\includegraphics[width=0.9\linewidth]{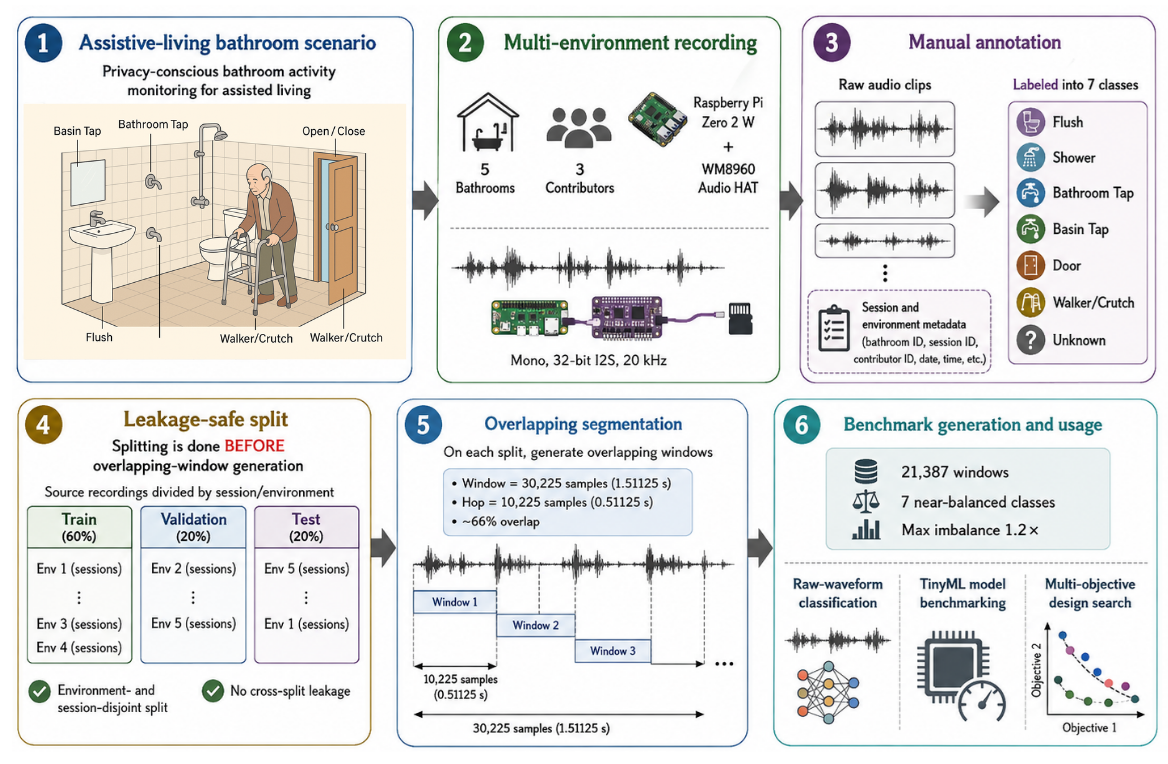}
	\caption{Overview of the \dataset{} data collection, annotation,
		preprocessing, and benchmark-generation pipeline.}
	\label{fig:pipeline}
\end{figure}

Source recordings are partitioned by recording session and environment
\emph{before} overlapping windows are generated. Consequently, windows that
share audio content or recording-specific characteristics remain within the
same training, validation, or test partition. A random window-level split
could otherwise place overlapping or closely related windows across different
partitions and allow room-specific cues to contribute to the reported
performance. Session- and environment-disjoint splitting is therefore used
throughout the study (Section~\ref{sec:setup}).
Figure~\ref{fig:pipeline} summarizes the data collection, annotation,
preprocessing, and partitioning procedure.

\subsection{Dataset Statistics}

\begin{figure}[!htbp]
	\centering
	\includegraphics[width=0.92\linewidth]{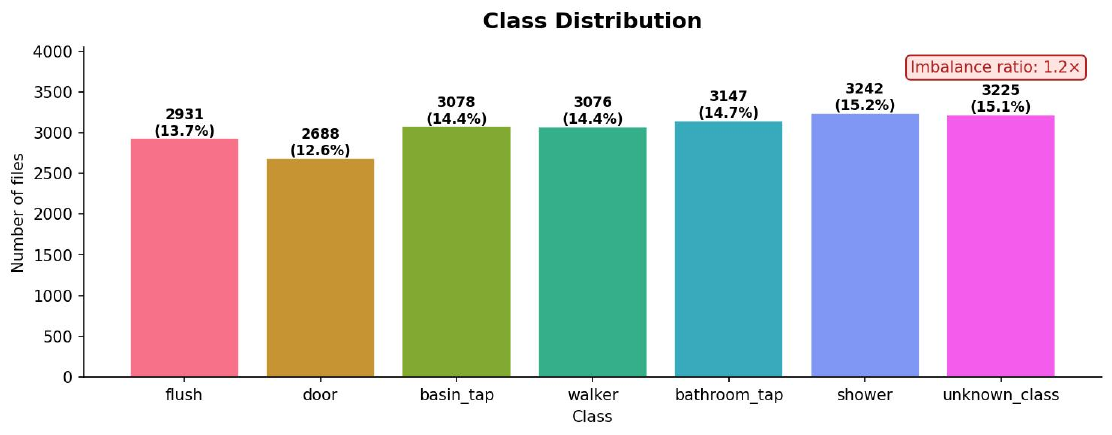}
	\caption{Class distribution of the \dataset{} dataset.}
	\label{fig:classdist}
\end{figure}

The corpus contains $21{,}387$ segmented windows with a near-balanced class
distribution (Fig.~\ref{fig:classdist}). The largest class
(\textit{Shower}, $3{,}242$) and the smallest
(\textit{Door}, $2{,}688$) give a maximum imbalance ratio of
$1.2\times$. Class imbalance is therefore limited in the present benchmark. Having established the corpus composition and partitioning protocol,
Section~\ref{sec:characterization} next examines the acoustic structure of
the seven classes before model training.

\section{Dataset Characterization}
\label{sec:characterization}

With the recording and preprocessing protocol established, we next examine
the corpus at the signal level. The aim is to identify acoustic similarities
and differences among the classes before any model is trained, and then test
whether these patterns are reflected in the learned representation and error
structure reported in Section~\ref{sec:results}. All analyses in this section
use hand-crafted signal descriptors and are therefore independent of the
learned model.

\subsection{Signal-Level and Spectral Structure}

\begin{figure}[!htbp]
    \centering
    \includegraphics[width=0.9\linewidth]{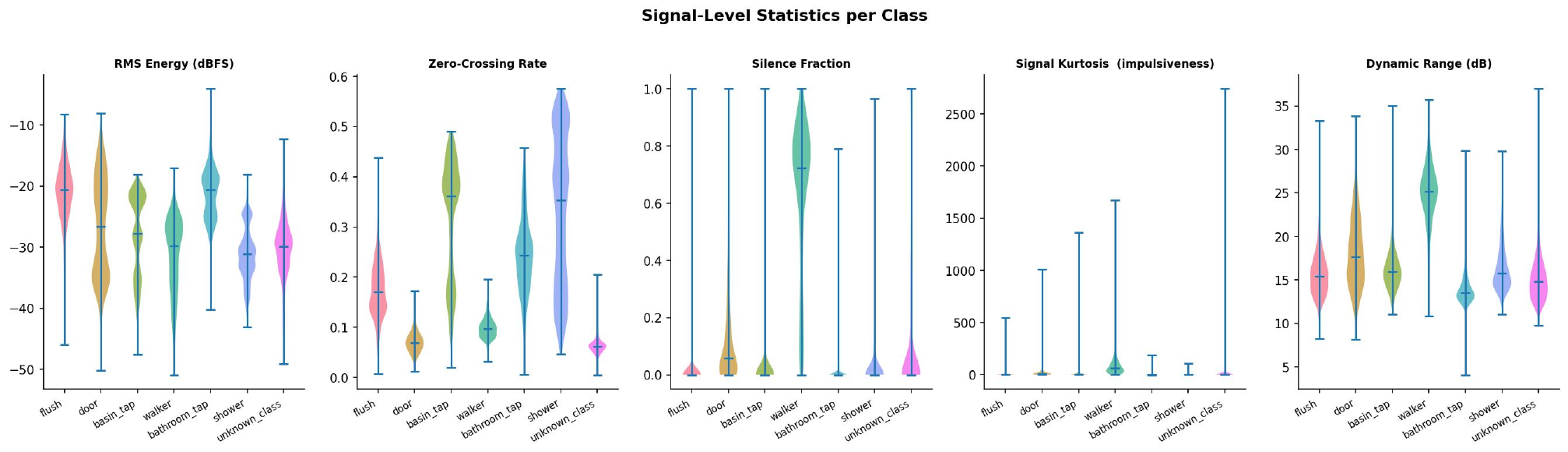}
    \caption{Signal-level statistics per class (violin plots). Impulsive
        classes (\textit{Door}, \textit{Walker/Crutch}) show markedly higher
        kurtosis and zero-crossing rate than the sustained water classes,
        providing a purely time-domain axis of separation that a raw-waveform
        model can exploit.}
    \label{fig:sigstats}
\end{figure}

We calculate low-level features such as the RMS energy, zero-crossing
rate (ZCR), spectral kurtosis, and silence ratio for each clip, along with the mean
power spectrum for each class. As shown in Fig.~\ref{fig:sigstats}, the classes are distinct on several of these measures. The \textit{Door} and
\textit{Walker/Crutch} clips, which include brief impulsive events,
usually have higher kurtosis values and temporally localized signals.
However, the \textit{Shower}, \textit{Bathroom Tap}, and
\textit{Flush} classes, which display extended broadband activity,
are at the other extreme, and the \textit{Basin Tap} class is intermediate.
Kruskal-Wallis tests show that there are differences among the seven classes
on RMS, ZCR, and kurtosis ($p<0.001$).

\begin{figure}[!htbp]
    \centering
    \includegraphics[width=0.6\linewidth]{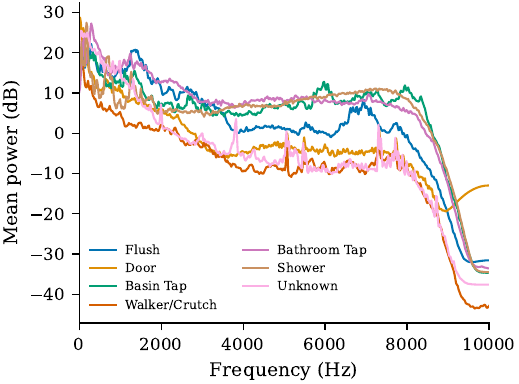}
    \caption{Per-class mean power spectra. The sustained water classes
        (\textit{Flush}, \textit{Shower}, \textit{Bathroom Tap}) overlap in the mid-frequency band, foreshadowing the model's principal confusions, whereas impulsive classes carry distinct high-frequency transient
        energy.}
    \label{fig:spectra}
\end{figure}

\begin{figure}[!htbp]
    \centering
    \includegraphics[width=01.0\linewidth]{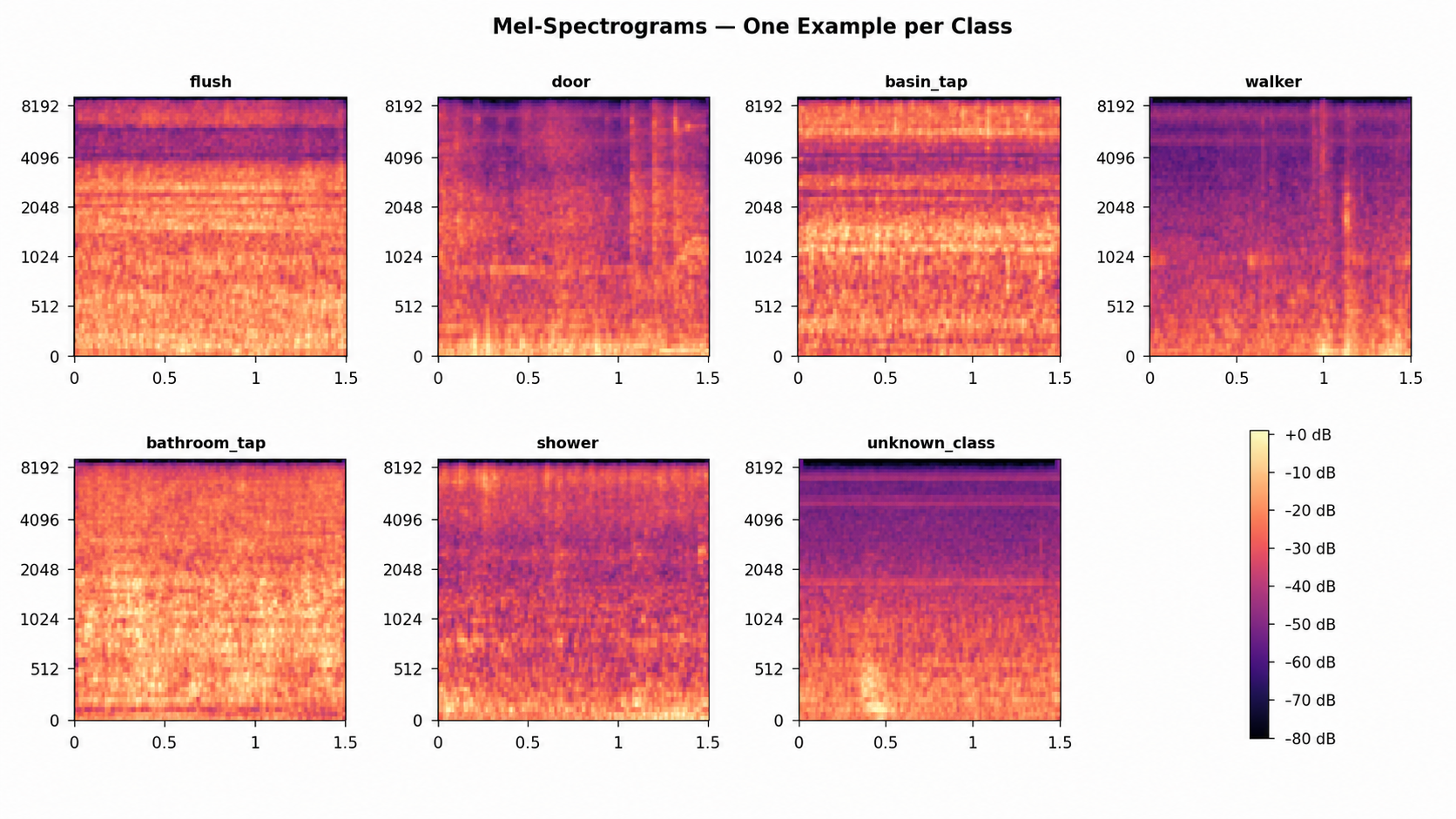}
    \caption{Representative mel-spectrogram per class. Impulsive events
        (\textit{Door}, \textit{Walker/Crutch}) appear as brief vertical onset
        stripes; the sustained water classes fill the window with visually
        similar broadband energy, previewing their confusability.}
    \label{fig:melspec}
\end{figure}

The power spectra provide a complementary view. As shown in
Fig.~\ref{fig:spectra}, the sustained water classes share substantial
mid-frequency energy, suggesting that spectral overlap may make some of
these classes difficult to separate. The mel-spectrogram examples in
Fig.~\ref{fig:melspec} show the same distinction in the time--frequency
domain. The impulsive events are dominated by short onset structures,
whereas \textit{Flush}, \textit{Bathroom Tap}, and \textit{Shower} contain
more sustained broadband energy.

\begin{figure}[!htbp]
    \centering
    \includegraphics[width=0.8\linewidth]{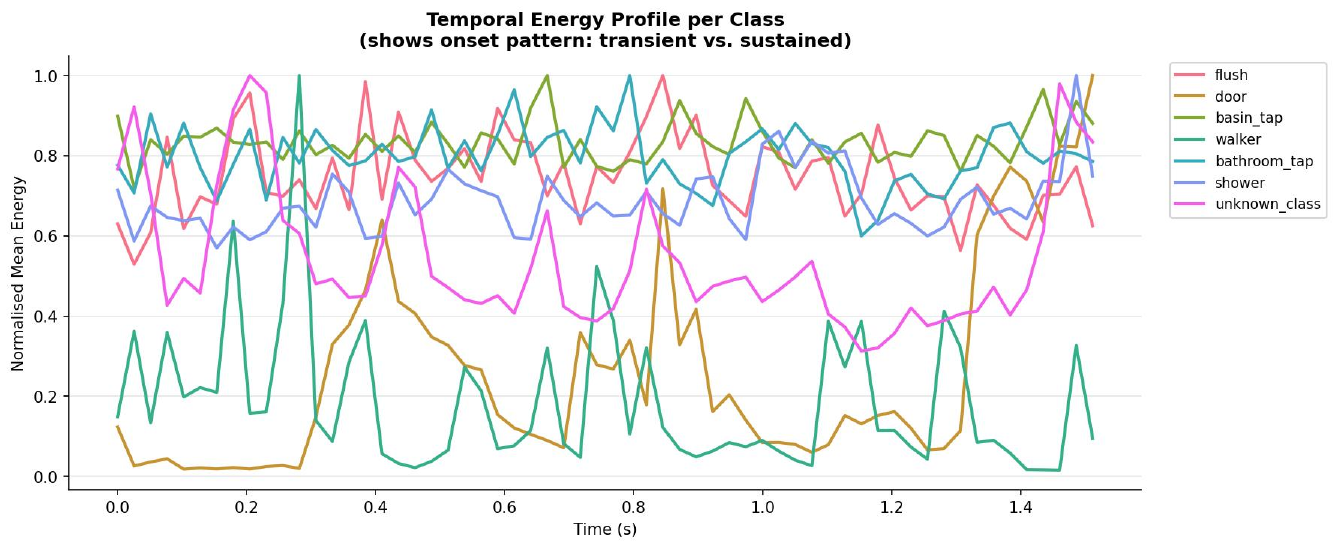}
    \caption{Per-class temporal energy profile. Impulsive events
        (\textit{Door}, \textit{Walker/Crutch}) show a sharp onset and decay;
        sustained water events maintain a flat envelope. The onset structure
        gives a time-domain cue that complements the overlapping spectra of
        Fig.~\ref{fig:spectra}.}
    \label{fig:temporal}
\end{figure}

Because SincDPNet operates directly on the waveform, temporal structure is
also relevant. Figure~\ref{fig:temporal} shows the average temporal energy
profile for each class. The impulsive events contain a sharp onset followed
by rapid decay, whereas the water-related classes maintain a more sustained
energy envelope. These differences provide temporal cues that complement the
spectral information in Fig.~\ref{fig:spectra}. Their usefulness to the
classifier is examined later through the learned representation and confusion
analysis.

\begin{figure}[!htbp]
    \centering
    \includegraphics[width=\linewidth]{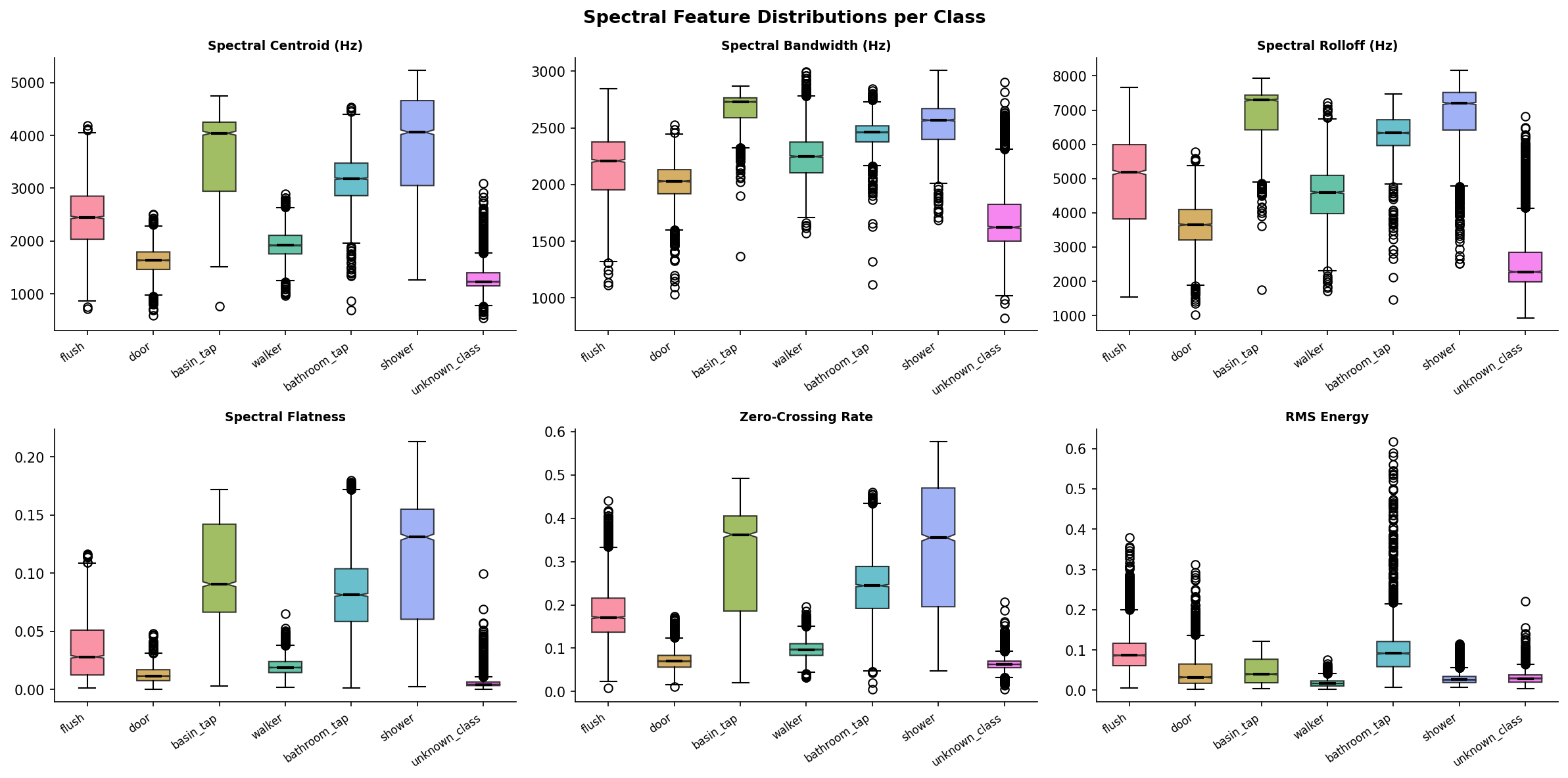}
    \caption{Per-class spectral descriptors (centroid, bandwidth, roll-off,
        flatness). Impulsive classes show higher centroids and flatness;
        sustained water classes cluster at lower centroids with smoother
        spectra, separating the two families along the same axis the model
        exploits.}
    \label{fig:spectralfeat}
\end{figure}

Figure~\ref{fig:spectralfeat} summarizes additional spectral descriptors,
including centroid, bandwidth, roll-off, and flatness. The impulsive classes
generally occupy higher-centroid and higher-flatness regions, while the
sustained water classes are concentrated at lower centroids with smoother
spectral profiles. These descriptors therefore provide another view of the
same broad acoustic separation.

\begin{figure}[!htbp]
    \centering
    \includegraphics[width=0.92\linewidth]{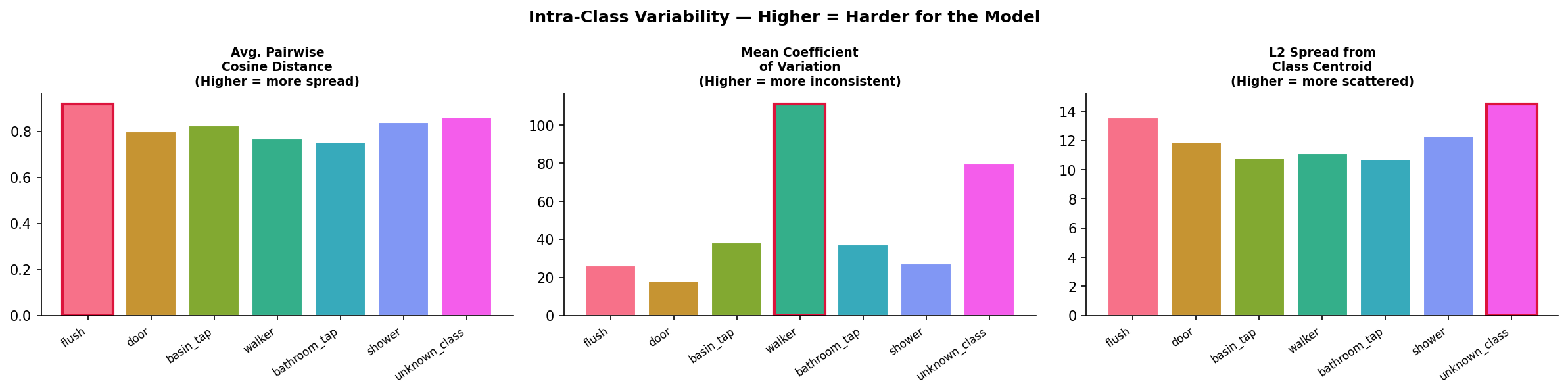}
    \caption{Within-class variability per class. The sustained water classes
        are internally variable (flow rate, basin geometry) as well as mutually
        overlapping, whereas impulsive classes are tighter. The combination of
        high intra-class spread and high inter-class overlap that makes the
        water triad the hardest group.}
    \label{fig:intraclass}
\end{figure}

Within-class variability is shown in Fig.~\ref{fig:intraclass}. The
water-related classes exhibit relatively large internal spread, which is
consistent with changes in flow rate, source position, and basin or room
geometry. This variability is important because a class may be difficult to
recognize not only when it overlaps with another class, but also when its own
acoustic distribution is broad.

\subsection{Inter-Class Overlap and a Difficulty Score}
The preceding analysis describes individual signal properties. We next
quantify how strongly the class distributions overlap. For each pair of
classes, we compute a Bhattacharyya-coefficient overlap using Gaussian
approximations of selected spectral features. For two class distributions
with means $\mu_i,\mu_j$ and variances $\sigma_i^2,\sigma_j^2$, the
Bhattacharyya coefficient is

\begin{equation}
    \label{eq:bc}
    \mathrm{BC}(i,j) = \exp\!\Big(-\tfrac{1}{4}\ln\!\Big[\tfrac{1}{4}\Big(\tfrac{\sigma_i^2}{\sigma_j^2}+\tfrac{\sigma_j^2}{\sigma_i^2}+2\Big)\Big] - \tfrac{1}{4}\tfrac{(\mu_i-\mu_j)^2}{\sigma_i^2+\sigma_j^2}\Big),
\end{equation}
with $\mathrm{BC}\in[0,1]$, where larger values indicate greater overlap.
The coefficient is averaged across the selected discriminative spectral
features.

To summarize several sources of difficulty in a single quantity, we combine
inter-class overlap, intra-class spread, class imbalance, silence fraction,
and energy variability into a per-class diagnostic score $D_c$,
\begin{equation}
    \label{eq:difficulty}
    D_c = \sum_{k} w_k \,\tilde{s}_{k,c},
\end{equation}
where $\tilde{s}_{k,c}\in[0,1]$ is the min--max-normalized value of diagnostic
$k$ for class $c$, $w_k\geq0$, and $\sum_k w_k=1$. Larger values therefore
indicate classes that appear more difficult according to the selected
signal-level diagnostics. We treat $D_c$ as a descriptive heuristic rather
than a predictor of model error; its sensitivity to the choice of weights is
examined in Section~\ref{sec:bridge}. Table~\ref{tab:difficulty} reports the
resulting ranking.

\begin{table}[H]
    \centering
    \caption{Per-class difficulty ranking from the signal-level analysis
        (higher $=$ predicted harder), computed before model training. The
        aggregate score is a heuristic: its class-level correlation with model
        error is weak ($\rho=-0.07$), while pairwise overlap is more useful for
        explaining specific confusions (Section~\ref{sec:bridge}).}
    \label{tab:difficulty}
    \renewcommand{\arraystretch}{1.1}
    \begin{tabular}{lccc}
        \toprule
        \textbf{Class} & \textbf{$D_c$} & \textbf{Mean overlap} & \textbf{Intra spread} \\
        \midrule
        Flush          & 0.74 & 0.50 & 0.92 \\
        Door           & 0.54 & 0.42 & 0.79 \\
        Basin Tap      & 0.45 & 0.46 & 0.81 \\
        Shower         & 0.43 & 0.47 & 0.83 \\
        Walker/Crutch  & 0.31 & 0.38 & 0.77 \\
        Bathroom Tap   & 0.25 & 0.42 & 0.76 \\
        Unknown Class  & 0.22 & 0.31 & 0.85 \\
        \bottomrule
    \end{tabular}
\end{table}

\begin{figure}[!htbp]
    \centering
    \includegraphics[width=\linewidth]{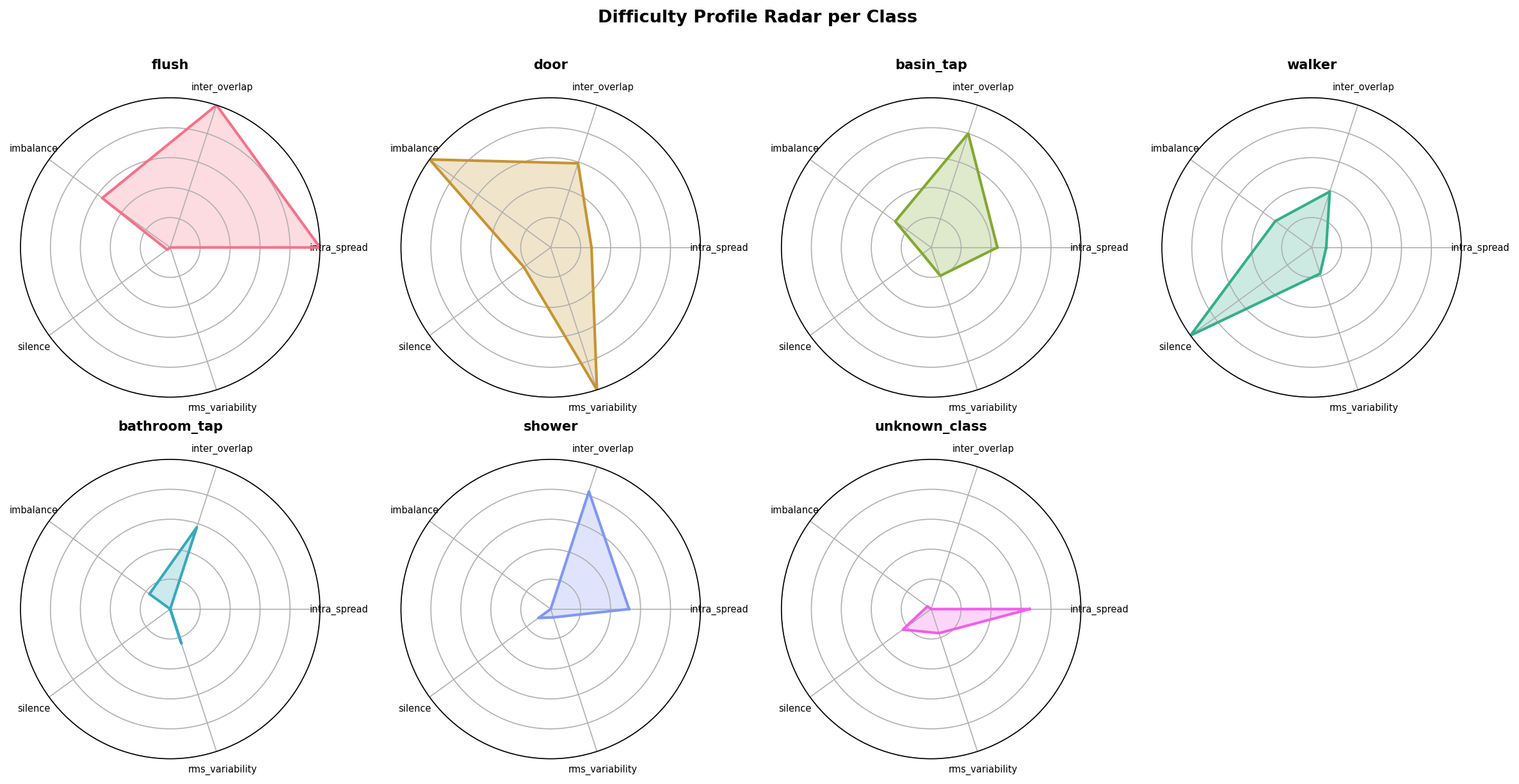}
    \caption{Per-class difficulty profiles across the five component
        diagnostics of Eq.~\eqref{eq:difficulty} (inter-class overlap,
        intra-class spread, class imbalance, silence fraction, energy
        variability). Each class is hard for a different combination of reasons;
        \textit{Flush} scores high on both overlap and spread, while the
        impulsive classes are driven by silence and energy variability.}
    \label{fig:radar}
\end{figure}

Figure~\ref{fig:radar} separates $D_c$ into its five component diagnostics.
The figure is useful because similar aggregate scores can arise for different
reasons. \textit{Flush}, for example, has relatively high inter-class overlap
and intra-class spread, whereas the impulsive classes are influenced more by
silence fraction and energy variability. The \textit{Unknown} class shows
substantial intra-class variation, consistent with its heterogeneous
composition.

\begin{figure}[!htbp]
    \centering
    \includegraphics[width=0.6\linewidth]{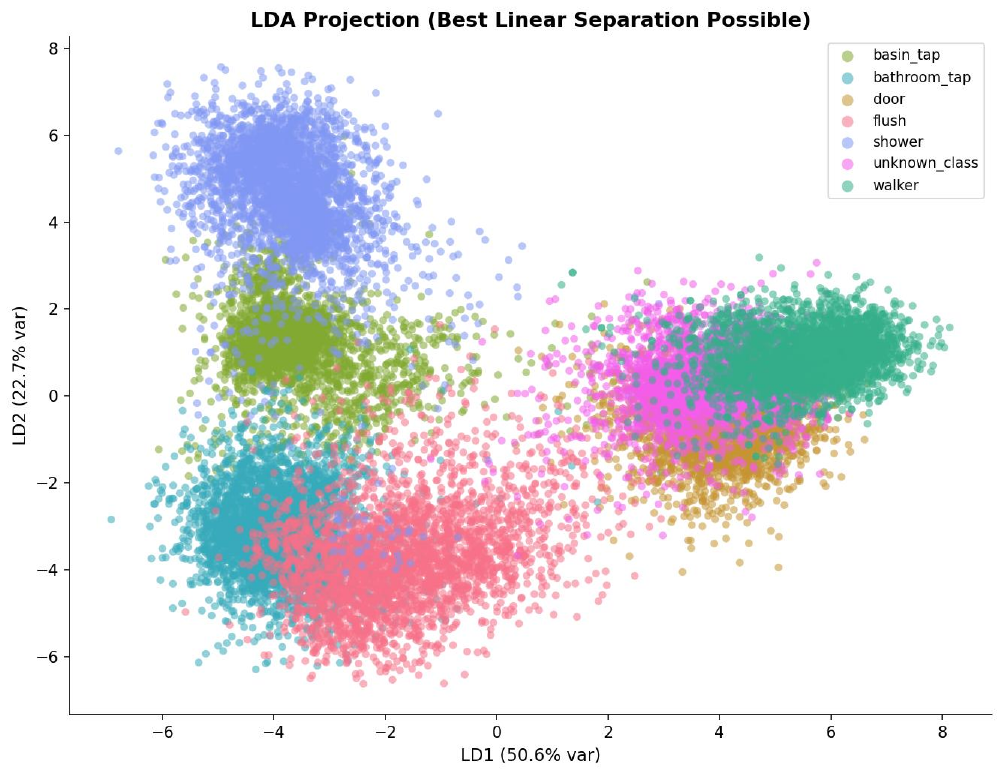}
    \caption{Linear-discriminant projection of the hand-crafted feature
        space. Impulsive classes and \textit{Basin Tap} separate cleanly, while
        the sustained water classes overlap even under the best linear
        projection, an intrinsic, model-independent property of the acoustics.}
    \label{fig:lda}
\end{figure}

A linear-discriminant projection provides an additional view of class
separability. Figure~\ref{fig:lda} projects the hand-crafted feature space
onto the first two discriminant axes. The impulsive classes and
\textit{Basin Tap} occupy relatively distinct regions, whereas
\textit{Flush}, \textit{Bathroom Tap}, and \textit{Shower} remain partially
overlapping. This pattern is consistent with the pairwise spectral analysis,
although the two-dimensional projection should be interpreted qualitatively.
A similar arrangement is later observed in the learned embedding space
(Fig.~\ref{fig:tsne_selected}).

\subsection{Design Implications}
The signal-level analysis provides two hypotheses for the model evaluation.
First, the sustained water classes are expected to produce substantial
cross-class confusion because of their spectral overlap and internal
variability. Second, successful discrimination should depend on cues that
remain distinct despite this overlap, including temporal onset structure and
frequency regions in which the class spectra differ.

These are hypotheses about the acoustic structure of the data rather than
claims about the behavior of a particular network. The extent to which they
are supported, contradicted, or refined by SincDPNet is examined later through
the learned filter responses, embedding structure, and confusion analysis in
Section~\ref{sec:bridge}.

\section{SincDPNet Architecture}
\label{sec:arch}

\begin{figure}[!htbp]
	\centering
	\includegraphics[width=\linewidth]{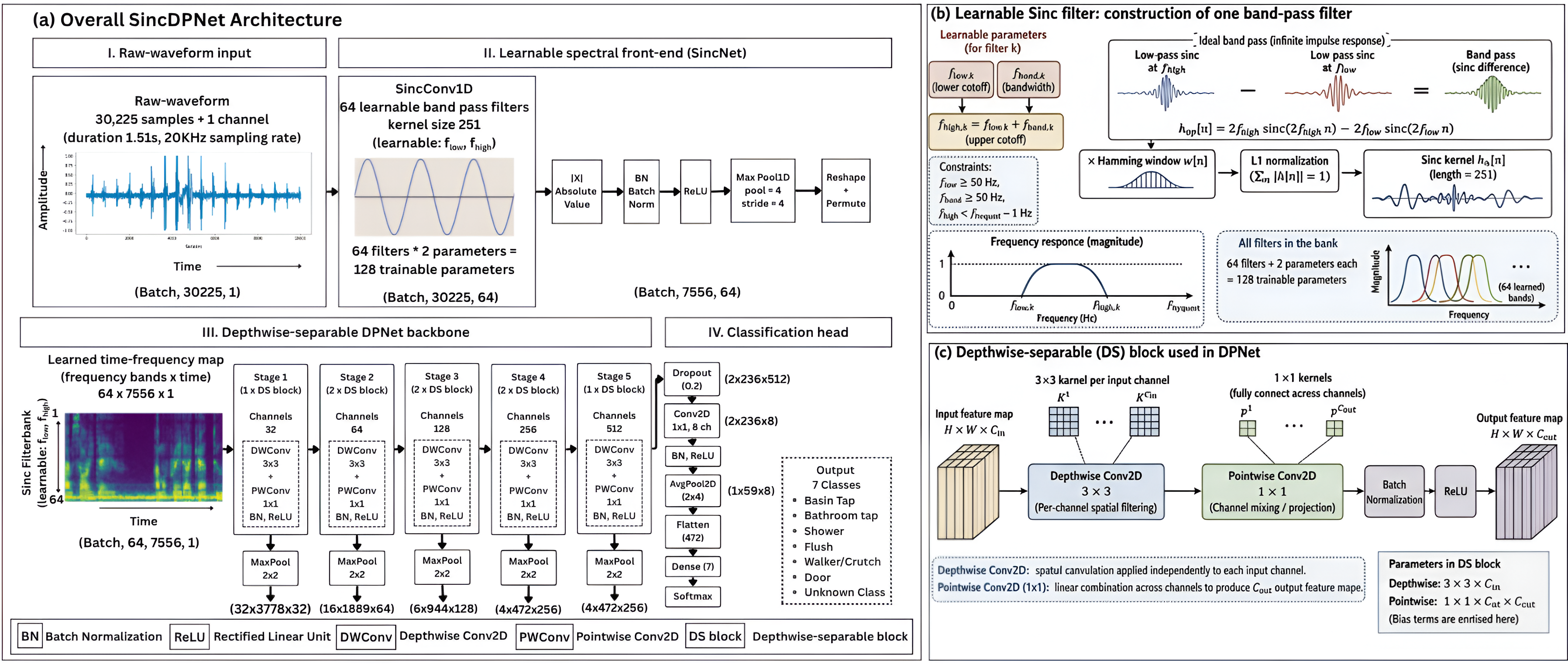}
	\caption{SincDPNet architecture. A learnable sinc filter bank,
		parameterized by per-filter cutoff and bandwidth in Hz, maps the raw
		waveform to $N_f$ band-pass channels. A depthwise-separable
		convolutional body then performs classification. The front end
		contains only $2N_f$ trainable parameters.}
	\label{fig:arch}
\end{figure}

The acoustic structure identified in Section~\ref{sec:characterization}
motivates a front end that is both frequency-aware and compact. SincDPNet
operates directly on the raw waveform and consists of two stages: a learnable
sinc band-pass front end that produces an interpretable time--frequency
representation, followed by a depthwise-separable convolutional body for
classification. Figure~\ref{fig:arch} shows the overall architecture.

\subsection{Learnable Sinc Band-Pass Front End}

The first layer is implemented as a filter bank whose filters are described by
frequency parameters rather than free convolutional weights. Consider the
ideal frequency response of a band-pass filter that passes frequencies
between $f_1$ and $f_2$:

\begin{equation}
	\label{eq:idealresp}
	G(f) = \mathrm{rect}\!\left(\frac{f}{2f_2}\right) -
	\mathrm{rect}\!\left(\frac{f}{2f_1}\right),
\end{equation}

where $\mathrm{rect}(\cdot)$ is $1$ on
$[-\tfrac12,\tfrac12]$ and $0$ elsewhere. Since the inverse Fourier
transform of a rectangle of width $2f_c$ is
$2f_c\,\mathrm{sinc}(2\pi f_c n)$, the corresponding time-domain
band-pass impulse response can be written as the difference of two
low-pass responses,

\begin{equation}
	\label{eq:bandpass}
	g[n; f_1,f_2] =
	2 f_2\,\mathrm{sinc}(2\pi f_2 n) -
	2 f_1\,\mathrm{sinc}(2\pi f_1 n),
\end{equation}

where $\mathrm{sinc}(x)=\sin(x)/x$, $\mathrm{sinc}(0)=1$, and
frequencies are normalized by the sampling rate $f_s$. Thus, a filter of
length $L$ is determined by only two frequency parameters rather than
$L$ independent coefficients. Following SincNet~\citep{ravanelli2018speaker},
the filter shape is fixed by Eq.~\eqref{eq:bandpass}, while its frequency
limits are learned from the data.

Filter $i$ is parameterized by a low cutoff $f_1^{(i)}$ and bandwidth
$b^{(i)}$:

\begin{equation}
	\label{eq:constraints}
	f_1^{(i)} = f_{\min} + |\hat f_1^{(i)}|, \qquad
	f_2^{(i)} =
	\min\!\big(f_1^{(i)} + f_{\min} + |\hat b^{(i)}|,\; f_s/2\big),
\end{equation}

so that the cutoffs remain positive and below the Nyquist frequency during
training. Only $\hat f_1^{(i)}$ and $\hat b^{(i)}$ are trainable. To reduce
spectral leakage caused by the finite kernel length $L$, each filter is
multiplied by a Hamming window $w[n]$,

\begin{equation}
	\label{eq:windowed}
	g_w[n;\,i] =
	g[n; f_1^{(i)}, f_2^{(i)}]\, w[n],
	\qquad
	n=-\tfrac{L-1}{2},\dots,\tfrac{L-1}{2}.
\end{equation}

The input waveform is convolved with the $N_f$ filters, followed by
magnitude computation, batch normalization, ReLU activation, and temporal
max-pooling. Importantly, the number of trainable front-end parameters is
independent of $L$: the complete filter bank requires only $2N_f$
parameters. Since these parameters define frequency limits, the learned
passbands can also be inspected directly in hertz after training
(Section~\ref{sec:filters}).

\paragraph{Relation to SincNet.}
SincDPNet adopts the sinc-filter parameterization introduced in
SincNet~\citep{ravanelli2018speaker}, in which each first-layer band-pass
filter is controlled by two frequency parameters rather than $L$ free
coefficients. The main architectural difference lies in the processing that
follows this front end. SincNet was developed for speaker recognition and
uses standard convolutional layers followed by large fully connected
classification layers, resulting in a model with on the order of
$10^5$ parameters.

SincDPNet instead reshapes the $N_f$ band-pass outputs into a
two-dimensional time--frequency representation
(Section~\ref{sec:body}) and processes it using a depthwise-separable
convolutional body. This substantially reduces the cost of the back end while
retaining the frequency interpretation of the sinc filters. The resulting
compact configuration contains only a few thousand parameters and is
approximately $28\times$ smaller than the SincNet implementation considered
in Table~\ref{tab:benchmark}. The architectural contribution therefore lies
in combining the interpretable sinc front end with a lightweight
depthwise-separable body designed for resource-constrained acoustic
classification.

\subsection{Depthwise-Separable Convolutional Body}
\label{sec:body}

After temporal pooling, the front end produces $N_f$ band-pass channels of
length $T$. These are reshaped into a two-dimensional map
$\mathbf{X}\in\mathbb{R}^{N_f\times T\times 1}$, where the vertical axis
indexes the learned frequency bands and the horizontal axis represents time.
The resulting representation is analogous to a spectrogram, except that the
frequency bands are learned directly from the waveform.

To keep the subsequent processing compact, the network uses
depthwise-separable convolutions. Consider a convolutional layer with a
$k\times k$ kernel that maps $C_\text{in}$ input channels to
$C_\text{out}$ output channels over a feature map of size $H\times W$.
A standard convolution requires

\begin{equation}
	\label{eq:stdconv}
	P_\text{std} = k^2\,C_\text{in}\,C_\text{out}
	\quad\text{parameters, and}\quad
	M_\text{std} =
	k^2\,C_\text{in}\,C_\text{out}\,H\,W
\end{equation}

multiply--accumulate operations. A depthwise-separable convolution
decomposes this operation into a depthwise convolution followed by a
pointwise convolution. The depthwise stage applies one $k\times k$ kernel
to each input channel independently, while the $1\times1$ pointwise stage
combines information across channels. The resulting costs are

\begin{equation}
	\label{eq:dsconv}
	P_\text{ds} =
	\underbrace{k^2 C_\text{in}}_{\text{depthwise}}
	+ \underbrace{C_\text{in} C_\text{out}}_{\text{pointwise}},
	\qquad
	M_\text{ds} =
	\big(k^2 C_\text{in} + C_\text{in} C_\text{out}\big) H W .
\end{equation}

Dividing Eq.~\eqref{eq:dsconv} by Eq.~\eqref{eq:stdconv} gives

\begin{equation}
	\label{eq:dsratio}
	\frac{P_\text{ds}}{P_\text{std}}
	=
	\frac{k^2 C_\text{in} + C_\text{in}C_\text{out}}
	{k^2 C_\text{in}C_\text{out}}
	=
	\frac{1}{C_\text{out}} + \frac{1}{k^2}.
\end{equation}

For the $k=3$ kernels used here, this becomes
$1/C_\text{out}+1/9$. As $C_\text{out}$ increases, the parameter cost
therefore approaches approximately one ninth of that of a corresponding
standard convolution.

The convolutional body contains $B$ such blocks. Each block applies a
depthwise-separable convolution, batch normalization, ReLU activation, and,
when the relevant spatial dimensions permit, $2\times2$ max-pooling. The
channel width is defined as

\begin{equation}
	\label{eq:width}
	C_b = \min\!\big(\alpha\,2^{\,b},\; 8\alpha\big),
	\qquad b = 0,\dots,B-1,
\end{equation}

where the width multiplier $\alpha$ controls the number of channels and the
growth saturates at $8\alpha$. The parameters $\alpha$ and $B$ therefore
control the size of the convolutional body and are included in the design
space explored in Section~\ref{sec:mobo}.

Global average pooling of the final feature map produces an embedding
$\mathbf{z}\in\mathbb{R}^{C_{B-1}}$. A linear classification layer followed
by softmax gives

\begin{equation}
	\label{eq:softmax}
	p(y=c\mid \mathbf{x}) =
	\frac{\exp(\mathbf{w}_c^{\!\top}\mathbf{z} + b_c)}
	{\sum_{c'=1}^{C}\exp(\mathbf{w}_{c'}^{\!\top}\mathbf{z} + b_{c'})},
\end{equation}

where $\mathbf{w}_c$ and $b_c$ denote the classifier weights and bias for
class $c$, and $C=7$ is the number of classes. The network is trained by
minimizing the cross-entropy loss

\[
\mathcal{L}
=
-\frac{1}{N}
\sum_{n=1}^{N}
\log p(y=y_n\mid\mathbf{x}_n),
\]

where $\mathbf{x}_n$ and $y_n$ denote the $n$th training clip and its
corresponding label.

\subsection{Parameter and Complexity Budget}

Table~\ref{tab:budget} summarizes the parameter allocation of the compact
$N_f=25$ configuration, one of the representative designs evaluated in
Section~\ref{sec:results}. The sinc filter bank contributes only a small
fraction of the total parameter count, while most parameters belong to the
convolutional body and classification head. Model size is therefore
controlled primarily by the body configuration and the number of sinc
filters, both of which are explored in the multi-objective design study of
Section~\ref{sec:mobo}.

\begin{table}[H]
	\centering
	\caption{Parameter budget of the compact SincDPNet ($N_f=25$).
		The filter-bank parameter count is independent of kernel length $L$.}
	\label{tab:budget}
	\renewcommand{\arraystretch}{1.1}
	\begin{tabular}{lcc}
		\toprule
		\textbf{Component} & \textbf{Parameters} & \textbf{Share} \\
		\midrule
		Sinc front-end ($2N_f$)         & 50      & 1.76\% \\
		Depthwise-separable body + head & 2{,}798 & 98.24\% \\
		\midrule
		\textbf{Total}                  & \textbf{2{,}848} & 100\% \\
		\bottomrule
	\end{tabular}
\end{table}
\section{Multi-Objective Design Optimization: A Practical Case Study}
\label{sec:mobo}

This section describes the design procedure used to select representative
SincDPNet configurations. A quality-based coreset is first used to reduce
the cost of evaluating candidate architectures. The design problem is then
studied through two bi-objective searches covering the accuracy--size and
accuracy--latency trade-offs. We finally examine the sensitivity of the
search to the acquisition function and surrogate kernel, and select
representative operating points from the resulting Pareto fronts.

\subsection{Scope and Framing}

Multi-objective Bayesian optimization (MOBO) is used here as an engineering
tool for exploring the SincDPNet design space rather than as a methodological
contribution. The aim is to make the model-selection procedure reproducible
while avoiding manual tuning of the accuracy--size and accuracy--latency
trade-offs. Section~\ref{sec:mobo-ablation} examines whether the selected
design region is sensitive to the acquisition function or surrogate kernel.

\begin{figure}[!htbp]
	\centering
	\begin{subfigure}[t]{0.65\linewidth}
		\centering
		\includegraphics[width=\linewidth]{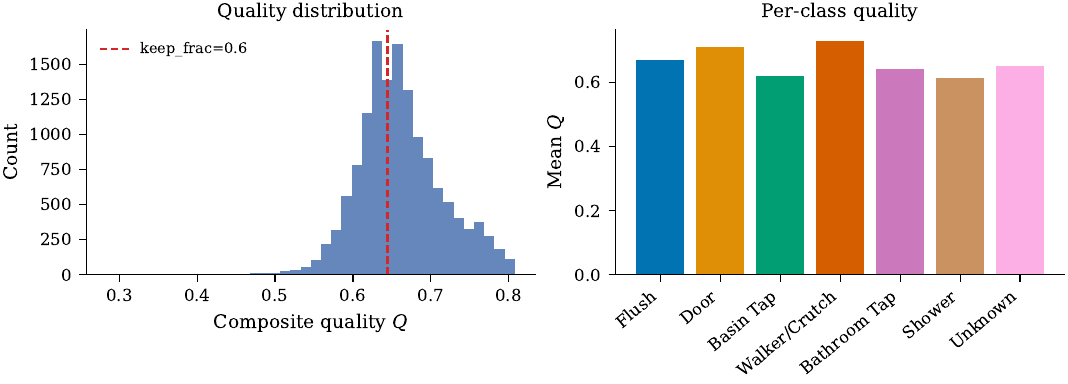}
		\caption{Quality-score distribution.}
		\label{fig:coreset}
	\end{subfigure}\hfill
	\begin{subfigure}[t]{0.33\linewidth}
		\centering
		\includegraphics[width=\linewidth]{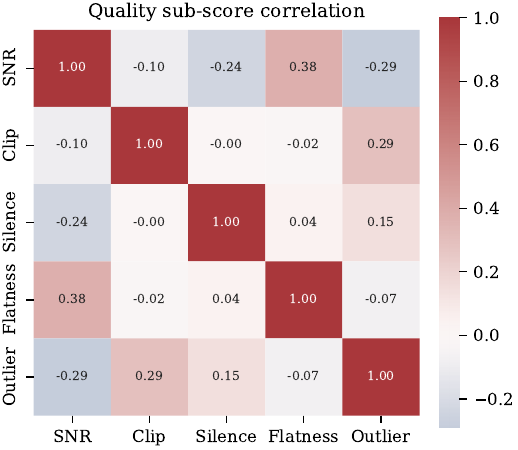}
		\caption{Coreset vs.\ full-data accuracy.}
		\label{fig:qualitycorr}
	\end{subfigure}
	\caption{Quality-based coreset construction. (a) Distribution of the
		composite quality score across training clips; the top $25\%$ per class
		is retained for search. (b) Validation macro-F1 for the same candidate
		architectures under coreset and full-data training. The reported rank
		correlation indicates how well the coreset preserves the ordering used
		during optimization.}
	\label{fig:coresetpair}
\end{figure}

To reduce the computational cost of each search evaluation, candidate models
are trained on a \emph{quality-based coreset}. Each training clip is assigned
a composite score based on estimated SNR, clipping fraction, silence
fraction, spectral flatness, and within-class outlierness. The highest-scoring
fraction within each class is retained for the search
(Fig.~\ref{fig:coresetpair}); in the experiments reported here, this fraction
is $25\%$.

The purpose of the coreset is to provide a lower-cost estimate of relative
candidate quality rather than to replace full-data training. Figure~\ref{fig:qualitycorr}
compares validation macro-F1 obtained from the $25\%$ coreset with the
corresponding result after full-data retraining for the same candidate
architectures. The observed rank agreement supports its use for ordering
candidates during search. All selected configurations are subsequently
retrained on the complete training set before final evaluation.

With the reduced evaluation set established, we next define the design
variables and optimization objectives.

\subsection{Search Space and Objectives}

Table~\ref{tab:searchspace} summarizes the design variables. Two independent
bi-objective studies are considered:

\begin{itemize}
	\item \textbf{Study A}: maximize validation macro-F1 and minimize model
	size (FP32 parameter footprint, KB);
	\item \textbf{Study B}: maximize validation accuracy and minimize
	inference latency (ms) measured on the target class of device.
\end{itemize}

The two studies are kept separate because they address different deployment
trade-offs and are influenced differently by the architectural variables.
Study~A emphasizes parameter footprint, whereas Study~B explicitly includes
measured inference time. Separate two-dimensional Pareto fronts also allow
the corresponding trade-offs to be inspected directly.

\begin{table}[H]
	\centering
	\caption{Design variables for the MOBO case study.}
	\label{tab:searchspace}
	\renewcommand{\arraystretch}{1.15}
	\begin{tabular}{lccp{0.30\linewidth}}
		\toprule
		\textbf{Variable} & \textbf{Symbol} & \textbf{Range} & \textbf{Primarily affects} \\
		\midrule
		Sinc filters       & $N_f$      & $16$--$128$          & size, F1, resolution \\
		Sinc kernel length & $L$        & $101$--$401$ (odd)   & latency, F1 \\
		Body width mult.\  & $\alpha$   & $2$--$12$            & size (dominant) \\
		DS-conv blocks     & $B$        & $3$--$6$             & size, latency \\
		Peak LR            & $\eta$     & $10^{-3}$--$3\!\times\!10^{-2}$ & F1 (convergence) \\
		Pooling stride     & $s$        & $\{2,4,8\}$          & latency, F1 \\
		\bottomrule
	\end{tabular}
\end{table}

\subsection{MOBO Formulation}

Let $\mathbf{x}$ denote a candidate configuration and
$\mathbf{f}(\mathbf{x})=(f_1(\mathbf{x}),f_2(\mathbf{x}))$ denote the two
objectives after sign adjustment so that both are minimized. An independent
Gaussian-process (GP) surrogate is fitted to each objective. Candidate
selection uses the $q$-Expected Hypervolume Improvement (qEHVI)
acquisition~\citep{daulton2020ehvi} relative to a reference point
$\mathbf{r}$ dominated by all feasible outcomes,

\begin{equation}
	\label{eq:ehvi}
	\mathbf{x}^\star = \arg\max_{\mathbf{x}} \;
	\mathbb{E}\!\left[\, \mathrm{HVI}\big(\{\mathbf{f}(\mathbf{x})\}\cup\mathcal{P}\big) \,\big|\, \mathcal{D} \right],
\end{equation}

where $\mathcal{P}$ is the current Pareto set, $\mathrm{HVI}$ denotes the
hypervolume improvement relative to $\mathbf{r}$, and $\mathcal{D}$ is the
set of evaluated configurations. The search begins with a Sobol design of
$n_0$ configurations and continues for a total budget of $T$ evaluations
\cite{Garai2026a}. Algorithm~\ref{alg:mobo} summarizes the procedure.

\begin{algorithm}[H]
	\caption{MOBO design loop (per study)}
	\label{alg:mobo}
	\begin{algorithmic}[1]
		\State Draw Sobol design $\{\mathbf{x}_1,\dots,\mathbf{x}_{n_0}\}$; evaluate $\mathbf{f}$ on \emph{validation} split
		\State Initialize Pareto set $\mathcal{P}$; fit GP surrogates
		\For{$t = n_0+1$ \textbf{to} $T$}
		\State $\mathbf{x}_t \gets \arg\max$ qEHVI (Eq.~\ref{eq:ehvi})
		\State Train configuration $\mathbf{x}_t$; evaluate $\mathbf{f}(\mathbf{x}_t)$ on validation
		\State Update $\mathcal{P}$; refit surrogates; log hypervolume
		\EndFor
		\State \Return Pareto set $\mathcal{P}$ and knee-point configurations
	\end{algorithmic}
\end{algorithm}

All objectives used during optimization are computed on the validation split.
The held-out test set is used only after configuration selection for the
results reported in Section~\ref{sec:results}. For Study~B, latency is
measured on the target device rather than inferred from training-GPU timing.

\subsection{Acquisition and Kernel Sensitivity}
\label{sec:mobo-ablation}

\begin{figure}[!htbp]
	\centering
	\includegraphics[width=0.6\linewidth]{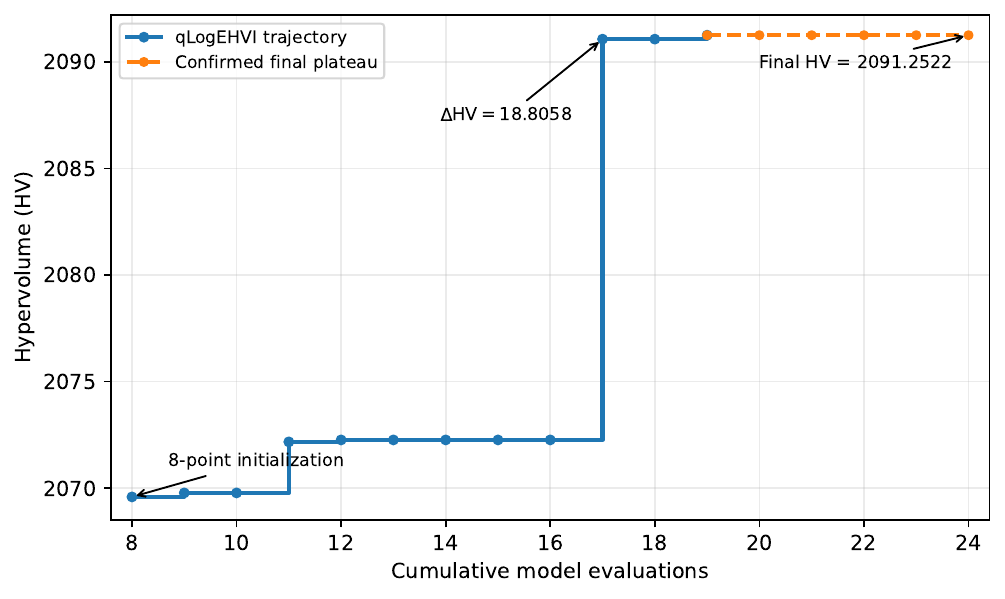}
	\caption{Hypervolume versus evaluation for Study~A. The surrogate-guided
		configurations reach similar final hypervolume values and outperform
		random search within the fixed evaluation budget.}
	\label{fig:hv}
\end{figure}

To examine whether the search outcome depends strongly on one optimizer
setting, Study~A is repeated with several acquisition functions and surrogate
kernels at the same evaluation budget. Table~\ref{tab:bo-ablation} compares
qEHVI, qNEHVI, and qParEGO with Mat\'ern-$5/2$ and RBF kernels, together with
a random/Sobol baseline. Figure~\ref{fig:hv} shows the corresponding
hypervolume trajectories.

The four surrogate-guided settings finish within $1.84$ hypervolume units of
one another, corresponding to approximately $0.088\%$ of the best final
value. In comparison, the difference between the best surrogate run
(C1, qEHVI with a Mat\'ern-$5/2$ kernel; HV $=2091.27$) and random search
(C5; HV $=2078.49$) is $12.78$ units, or approximately $0.61\%$.

The convergence rates show the same pattern. C1 reaches $90\%$ of its
hypervolume gain after 17 evaluations, and all surrogate-guided settings
reach this level within 22 evaluations. Random search does not reach the
same threshold within the 24-evaluation budget. Within the settings examined
here, the final design is therefore relatively insensitive to the particular
acquisition/kernel combination, while surrogate-guided search reaches the
high-hypervolume region more consistently within the available budget.

\begin{table}[H]
	\centering
	\footnotesize
	\setlength{\tabcolsep}{3.5pt}
	\caption{Acquisition and kernel sensitivity for Study~A at a fixed budget
		of 24 model evaluations. Higher final hypervolume (HV) is better,
		while ``Evals to 90\%'' indicates the evaluations required to reach
		90\% of the best HV gain. Best values are shown in \textbf{bold}.}
	\label{tab:bo-ablation}
	\renewcommand{\arraystretch}{1.15}
	\begin{tabular}{llccc}
		\toprule
		\textbf{Config} & \textbf{Acquisition} & \textbf{Kernel} &
		\textbf{Final HV} & \textbf{Evals to 90\%} \\
		\midrule
		C1 & qEHVI   & Mat\'ern-5/2 & \textbf{2091.27} & \textbf{17} \\
		C2 & qNEHVI  & Mat\'ern-5/2 & 2090.65 & 18 \\
		C3 & qParEGO & Mat\'ern-5/2 & 2089.43 & 22 \\
		C4 & qEHVI   & RBF          & 2090.12 & 19 \\
		C5 & Random  & ---          & 2078.49 & $>24$ \\
		\bottomrule
	\end{tabular}
\end{table}

\subsection{Configuration Selection by Weighted Scoring}

\begin{figure}[!htbp]
	\centering
	\begin{subfigure}[t]{0.38\linewidth}
		\centering
		\includegraphics[width=\linewidth]{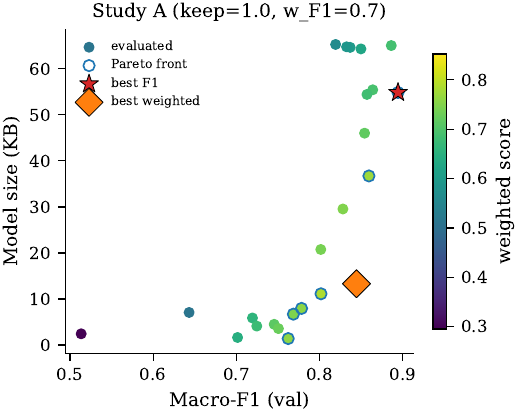}
		\caption{Study A: F1 vs.\ size.}
		\label{fig:pareto-a}
	\end{subfigure}\hfill
	\begin{subfigure}[t]{0.38\linewidth}
		\centering
		\includegraphics[width=\linewidth]{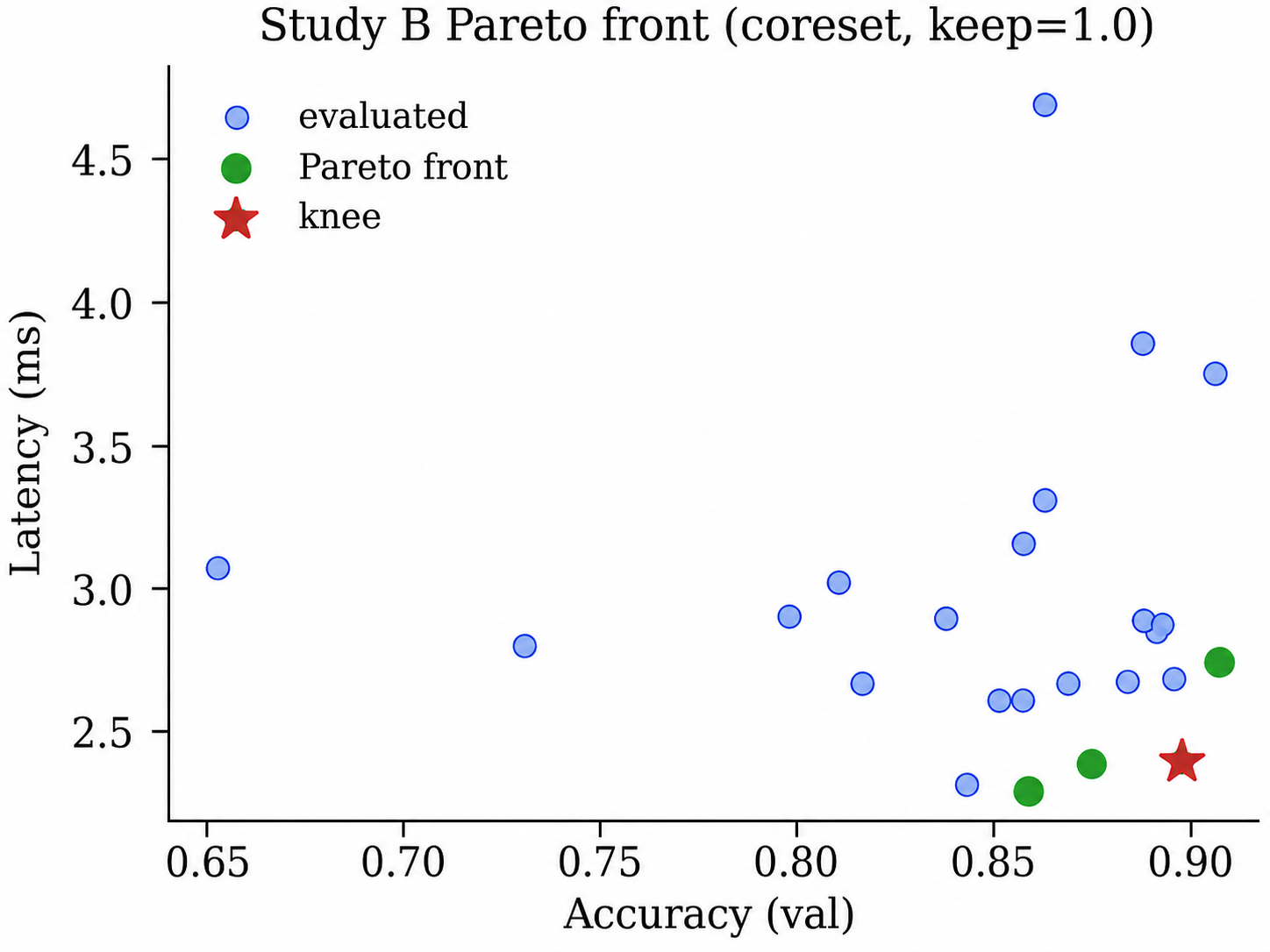}
		\caption{Study B: accuracy vs.\ latency.}
		\label{fig:pareto-b}
	\end{subfigure}
	\caption{Pareto fronts from the two design studies. Dominated
		configurations are shown faded; points are colored by weighted score,
		and the highest-F1 and highest weighted-score selections are marked.}
	\label{fig:pareto}
\end{figure}

Figure~\ref{fig:pareto} shows the Pareto fronts obtained from the two design
studies. For Study~A, a weighted score is also recorded for each of the
24 evaluated configurations to summarize the validation macro-F1--size
trade-off. The highest-scoring configuration uses $N_f=78$, $\alpha=4$, and
five blocks, with validation macro-F1 $0.8443$, model size $13.3$~KB, and a
weighted score of $0.8523$. The highest validation macro-F1, $0.8942$, is
obtained by the larger $N_f=59$, $\alpha=10$, five-block configuration,
which occupies $54.8$~KB. The two configurations therefore represent
different operating points on the same accuracy--size trade-off.

The non-dominated Study-A configurations cover a broad range of model sizes
and validation macro-F1 values, from $1.4$~KB and $0.7622$ macro-F1 to
$54.8$~KB and $0.8942$. Seven of the 24 configurations are Pareto-optimal in
the macro-F1--size plane. The results also show that a larger parameter
footprint does not necessarily produce higher validation performance; several
configurations near $65$~KB, for example, obtain lower macro-F1 than the
$13.3$~KB highest-score model. Table~\ref{tab:ranking} reports all evaluated
Study-A configurations. The compact $N_f=25$ configuration used elsewhere
in the paper is retained as one of the representative operating points.

\begin{table}[!t]
	\caption{Study-A configurations ranked according to the weighted score.
		$N_f$ denotes the number of sinc filters, $\alpha$ denotes the width
		multiplier, and $B$ denotes the number of depthwise-separable blocks.
		A $\star$ indicates a Pareto-optimal configuration in the
		Macro-F1--model-size plane. Bold values indicate the best overall values.}
	\label{tab:ranking}
	
	\centering
	\fontsize{7.5}{8.5}\selectfont
	\setlength{\tabcolsep}{3.0pt}
	\renewcommand{\arraystretch}{0.92}
	
	\begin{tabular*}{\textwidth}{@{\extracolsep{\fill}}cccccccc@{}}
		\toprule
		
		\textbf{Rank} &
		\multicolumn{3}{c}{\textbf{Architecture}} &
		\multicolumn{2}{c}{\textbf{Performance}} &
		\multicolumn{2}{c@{}}{\textbf{Selection}} \\
		
		\cmidrule(lr){2-4}
		\cmidrule(lr){5-6}
		\cmidrule(l){7-8}
		
		&
		$N_f$ & $\alpha$ & $B$ &
		Size (KB) & Macro-F1 &
		Weighted & Pareto \\
		
		\midrule
		
		1  & 78  & 4  & 5 & 13.3 & 0.8443 & \textbf{0.8523} & $\star$ \\
		2  & 25  & 6  & 4 & 11.1 & 0.8016 & 0.7842 & $\star$ \\
		3  & 24  & 8  & 5 & 36.7 & 0.8592 & 0.7698 & $\star$ \\
		4  & 20  & 2  & 3 & 1.4  & 0.7622 & 0.7574 & $\star$ \\
		5  & 24  & 3  & 5 & 7.9  & 0.7783 & 0.7562 & $\star$ \\
		6  & 59  & 10 & 5 & 54.8 & \textbf{0.8942} & 0.7490 & $\star$ \\
		7  & 34  & 7  & 5 & 29.5 & 0.8279 & 0.7460 & -- \\
		8  & 110 & 2  & 5 & 6.7  & 0.7684 & 0.7439 & $\star$ \\
		9  & 23  & 9  & 4 & 20.7 & 0.8014 & 0.7385 & -- \\
		10 & 62  & 4  & 3 & 3.6  & 0.7505 & 0.7256 & -- \\
		11 & 69  & 9  & 5 & 46.0 & 0.8540 & 0.7166 & -- \\
		12 & 16  & 2  & 5 & 4.5  & 0.7454 & 0.7119 & -- \\
		13 & 84  & 10 & 5 & 55.4 & 0.8638 & 0.6904 & -- \\
		14 & 72  & 11 & 5 & 65.0 & 0.8861 & 0.6862 & -- \\
		15 & 41  & 10 & 5 & 54.4 & 0.8568 & 0.6823 & -- \\
		16 & 112 & 3  & 3 & 4.1  & 0.7245 & 0.6753 & -- \\
		17 & 60  & 7  & 3 & 5.9  & 0.7192 & 0.6574 & -- \\
		18 & 30  & 2  & 3 & 1.6  & 0.7014 & 0.6445 & -- \\
		19 & 41  & 11 & 5 & 64.3 & 0.8496 & 0.6225 & -- \\
		20 & 55  & 11 & 5 & 64.6 & 0.8366 & 0.5972 & -- \\
		21 & 60  & 11 & 5 & 64.8 & 0.8324 & 0.5888 & -- \\
		22 & 82  & 11 & 5 & 65.3 & 0.8192 & 0.5621 & -- \\
		23 & 111 & 7  & 3 & 7.1  & 0.6431 & 0.5119 & -- \\
		24 & 64  & 2  & 3 & 2.4  & 0.5134 & 0.2952 & -- \\
		
		\bottomrule
	\end{tabular*}
	
\end{table}

The representative operating points selected from this analysis are evaluated
under the environment-disjoint protocol in Section~\ref{sec:results}.

\section{Experiments and Results}
\label{sec:results}
This section evaluates SincDPNet under environment-disjoint splitting. We
first describe the experimental protocol and compare the proposed models with
representative raw-waveform and MFCC-based baselines. We then examine the
embedding structure, architectural ablations, and learned sinc filters.
Finally, the observed classification errors are related to the acoustic
properties of the corresponding classes.
The source code, model implementation, and experimental scripts used in this
study are publicly available in the project
repository.\footnote{\url{https://github.com/debolina-34/SnaanGhar7}}
\subsection{Experimental Setup}
\label{sec:setup}
We compare SincDPNet against six baselines spanning both input domains:
raw-waveform models (DPNet~\citep{chowdhury2026dpnet},
ACDNet~\citep{mohaimenuzzaman2023environmental},
SincNet~\citep{ravanelli2018speaker}) and MFCC models
(DS-CNN~\citep{zhang2017hello},
CRNN~\citep{cakir2017convolutional}, TinyCNN~\citep{zhang2017hello}).

\paragraph{Splitting.}
The recordings come from five environments, each with its own room response.
A random clip-level split would therefore place acoustically related windows
from the same environment in both training and test sets. We avoid this by
using an environment-disjoint split: each recording environment is assigned
entirely to training, validation, or testing. All models are evaluated using
the same assignments. We additionally report leave-one-environment-out
(LOEO) cross-validation to assess generalization across environments

\paragraph{Fixed split and LOEO protocol.}
For the main experiment, E01--E03 are used for training, E04 for validation,
and E05 for testing. The LOEO evaluation contains five folds. In each fold,
one environment is reserved for testing, one of the remaining environments
is used for validation, and the other three are used for training. Model
selection uses only the validation environment.

%

\paragraph{Training and metrics.}
Waveforms are peak-normalized and center-cropped or padded to $30{,}225$
samples. Amplitude augmentation of $\pm25\%$ is applied only to the training
set after splitting. Models are trained using class-balanced categorical
cross-entropy. Accuracy, balanced accuracy, macro-F1, and Matthews
correlation coefficient (MCC) are reported as mean~$\pm$~standard deviation
over three random seeds. Macro-F1 and MCC are given particular attention
because they provide information about class-wise performance that is not
captured by overall accuracy alone.

\subsection{Benchmark Results}
Table~\ref{tab:benchmark} summarizes the environment-disjoint results. The
baseline models obtain accuracies between 69.3\% and 76.7\% and macro-F1
scores between 0.612 and 0.712. These values are substantially lower than the
near-saturated results obtained with random clip-level splitting on the same
corpus. The difference is expected because random splitting allows windows
from the same recording environment to occur in both training and test sets.
The environment-disjoint protocol therefore provides a more demanding
estimate of performance in a previously unseen room.

Among the proposed configurations, the best-F1 SincDPNet reaches 80.2\%
accuracy, 0.847 balanced accuracy, and 0.760 macro-F1 with 14,040
parameters. The rank-1 configuration reduces the model to 3,408 parameters
while retaining 0.673 macro-F1 and 0.714 MCC. The compact $N_f=25$ model is
the smallest model in the comparison, with 2,848 parameters, and obtains
75.7\% accuracy, 0.661 macro-F1, and 0.716 MCC.

The compact model is particularly informative when compared with DPNet,
since the two models have similar lightweight backbones. Replacing the
generic convolutional front end with the sinc parameterization increases
macro-F1 from 0.612 to 0.661 and MCC from 0.645 to 0.716 while reducing the
parameter count from 4,944 to 2,848. At the other end of the trade-off, the
best-F1 SincDPNet achieves the highest accuracy, balanced accuracy, and
macro-F1 in Table~\ref{tab:benchmark}, whereas TinyCNN retains the highest
MCC. The results therefore provide several useful operating points rather
than a single configuration that is preferable under every resource budget.


\begin{table}[!t]
    \centering
    \caption{Performance comparison on \dataset{} under environment-disjoint
        splitting. Results are reported as mean $\pm$ standard deviation over three
        random seeds. Parameter counts are exact. The best result in each metric is
        shown in bold.}
    \label{tab:benchmark}
    
    \small
    \setlength{\tabcolsep}{4pt}
    \renewcommand{\arraystretch}{1.15}
    
    \begin{tabular}{llrcccc}
        \toprule
        \textbf{Model} &
        \textbf{Input} &
        \textbf{Params} &
        \textbf{Acc. (\%)} &
        \textbf{Bal. Acc.} &
        \textbf{Macro-F1} &
        \textbf{MCC} \\
        \midrule
        
        DS-CNN~\citep{zhang2017hello}
        & MFCC & 18{,}311
        & \meanstd{75.3}{6.8}
        & \meanstd{0.811}{0.053}
        & \meanstd{0.705}{0.063}
        & \meanstd{0.715}{0.071} \\
        
        CRNN~\citep{cakir2017convolutional}
        & MFCC & 236{,}967
        & \meanstd{74.6}{4.3}
        & \textbf{\meanstd{0.826}{0.024}}
        & \meanstd{0.695}{0.027}
        & \meanstd{0.709}{0.041} \\
        
        TinyCNN~\citep{zhang2017hello}
        & MFCC & 93{,}575
        & \meanstd{76.5}{7.4}
        & \meanstd{0.809}{0.058}
        & \meanstd{0.698}{0.070}
        & \textbf{\meanstd{0.729}{0.076}} \\
        
        SincNet~\citep{ravanelli2018speaker}
        & Raw & 132{,}327
        & \textbf{\meanstd{76.7}{1.8}}
        & \meanstd{0.809}{0.027}
        & \textbf{\meanstd{0.712}{0.013}}
        & \meanstd{0.721}{0.023} \\
        
        ACDNet~\citep{mohaimenuzzaman2023environmental}
        & Raw & 22{,}055
        & \meanstd{71.7}{3.0}
        & \meanstd{0.729}{0.032}
        & \meanstd{0.639}{0.033}
        & \meanstd{0.670}{0.028} \\
        
        DPNet~\citep{chowdhury2026dpnet}
        & Raw & 4{,}944
        & \meanstd{69.3}{2.0}
        & \meanstd{0.689}{0.016}
        & \meanstd{0.612}{0.009}
        & \meanstd{0.645}{0.020} \\
        
        \midrule
        
        \multicolumn{7}{l}{\textit{Proposed SincDPNet configurations}} \\
        \midrule
        
        SincDPNet best-F1
        & Raw & 14{,}040
        & \meanstd{80.2}{3.2}
        & \meanstd{0.847}{0.022}
        & \meanstd{0.760}{0.019}
        & \meanstd{0.716}{0.052} \\
        
        SincDPNet rank-1
        & Raw & 3{,}408
        & \meanstd{75.6}{1.8}
        & \meanstd{0.821}{0.012}
        & \meanstd{0.673}{0.029}
        & \meanstd{0.714}{0.020} \\
        
        SincDPNet compact ($N_f$=25)
        & Raw & \textbf{2{,}848}
        & \meanstd{75.7}{3.8}
        & \meanstd{0.801}{0.062}
        & \meanstd{0.661}{0.049}
        & \meanstd{0.716}{0.042} \\
        
        \bottomrule
    \end{tabular}
    
    \vspace{2pt}
    \footnotesize
    \textit{Note:} All results are reported as mean $\pm$ standard deviation over
    three random seeds. Parameter counts are exact.
\end{table}

\subsection{Raw-Waveform versus Spectral-Domain Analysis}
The benchmark also allows us to examine whether raw-waveform input alone
provides an advantage over MFCC-based representations. Among the baseline
models, there is no consistent separation between the two input domains.
SincNet obtains a macro-F1 of 0.712, slightly above the 0.705 of DS-CNN,
whereas the raw-waveform ACDNet and DPNet obtain lower scores of 0.639 and
0.612, respectively.

These results suggest that the choice of front end is more consequential
than the input representation by itself. In particular, both SincNet and
SincDPNet constrain the first layer to learn band-pass filters, whereas
ACDNet and DPNet use generic convolutional filters. The direct comparison
between the compact SincDPNet and DPNet supports this interpretation:
SincDPNet achieves higher macro-F1 and MCC with fewer parameters. For the
present small-data setting, imposing physically meaningful structure on the
front end appears to be an effective use of a limited parameter budget. To see
whether this distinction is also reflected in the learned representation, we
next examine the test-set embedding space.

\subsection{Embedding-Space Structure}
\label{sec:embedding}
\begin{figure}[!htbp]
    \centering
    \begin{subfigure}[t]{0.39\textwidth}
        \centering
        \includegraphics[width=\linewidth]{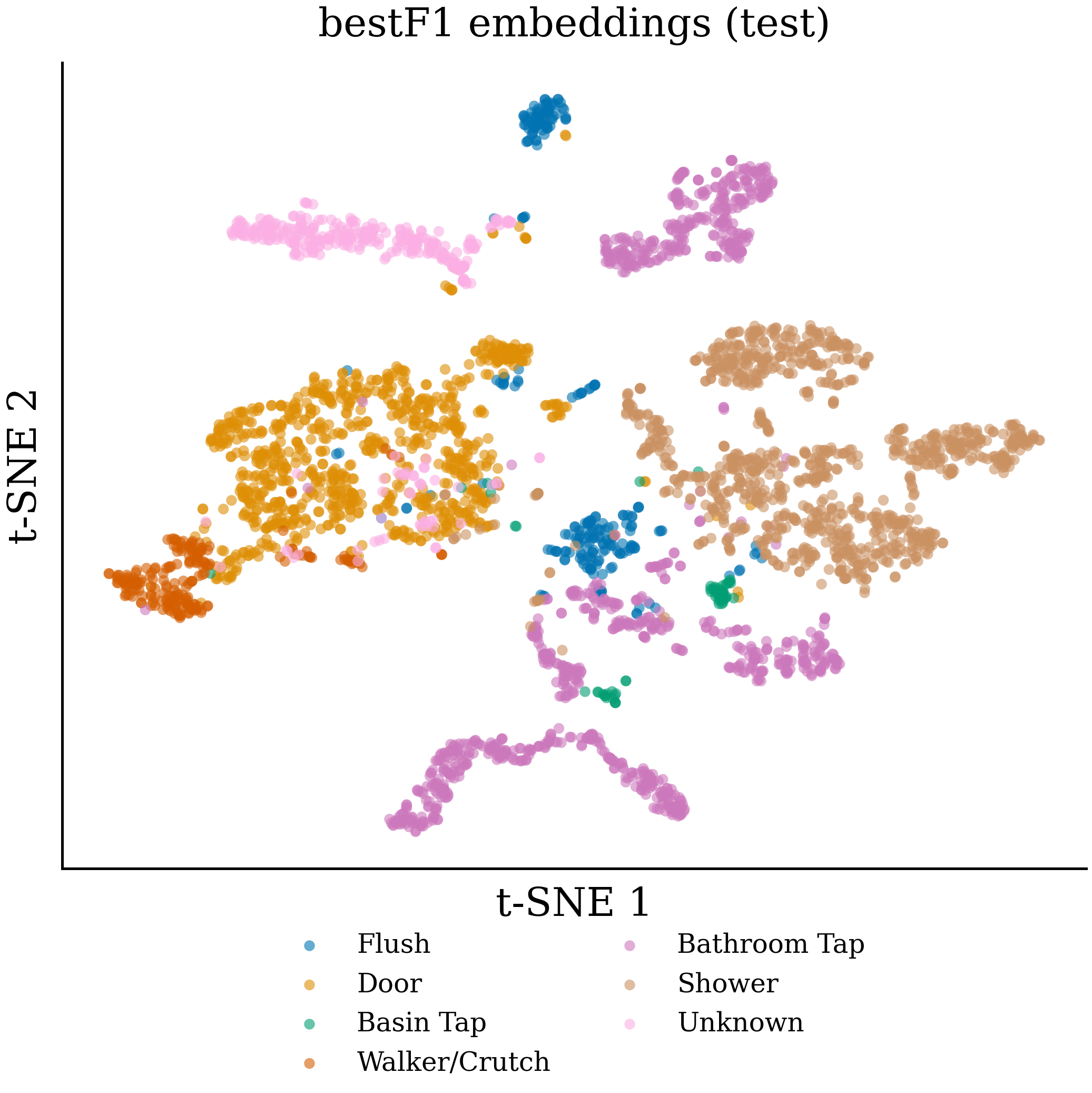}
        \caption{Best-macro-F1 model ($N_f=59$, $\alpha=10$, $5$ blocks).}
        \label{fig:tsne_bestf1}
    \end{subfigure}\hfill
    \begin{subfigure}[t]{0.39\textwidth}
        \centering
        \includegraphics[width=\linewidth]{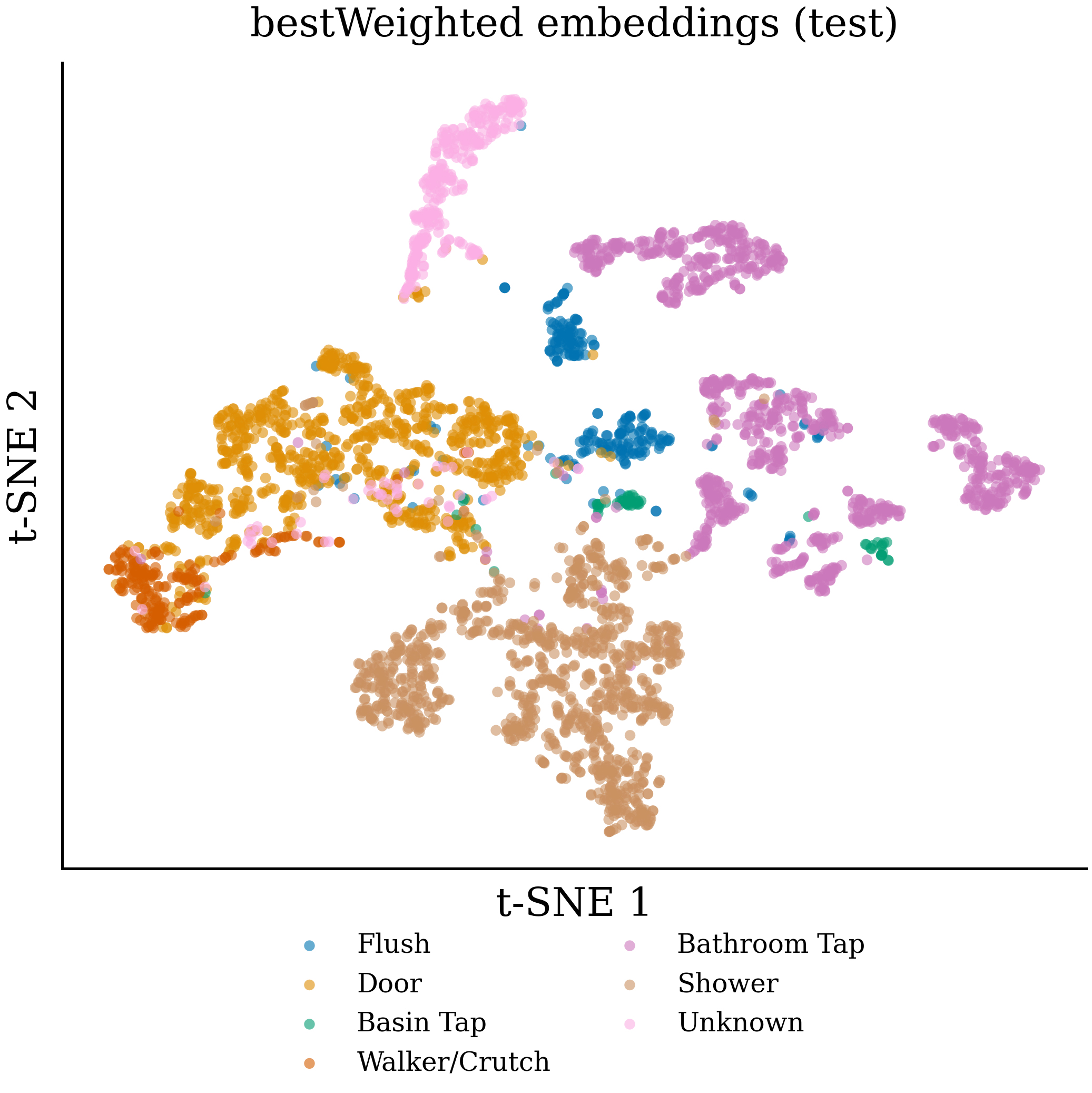}
        \caption{Rank-1 weighted-score model ($N_f=78$, $\alpha=4$, $5$ blocks).}
        \label{fig:tsne_bestw}
    \end{subfigure}
    \caption{Test-set t-SNE embeddings of the two configurations selected
        by the search. The best-F1 model (a) uses more capacity and pulls the
        water classes slightly further apart, while the compact rank-1 model
        (b) keeps the same overall layout at a fraction of the size. In both,
        the impulsive classes separate cleanly and the sustained water classes
        share one region, so the structure is preserved across the trade-off.}
    \label{fig:tsne_selected}
\end{figure}

Figure~\ref{fig:tsne_selected} shows two-dimensional t-SNE projections of
the penultimate-layer embeddings for the held-out test set. Since these embeddings are obtained from samples that were not
used for training, the plots provide a qualitative view of how the learned
representation behaves in the unseen environment.

Several patterns are visible. The sustained water classes
(\textit{Flush}, \textit{Bathroom Tap}, and \textit{Shower}) occupy adjacent
and partially overlapping regions, consistent with their spectral overlap.
The \textit{Unknown} samples are more dispersed, as expected for a class
containing heterogeneous non-target sounds. The impulsive classes and
\textit{Basin Tap} form comparatively compact regions in the two-dimensional
projection.

The last observation should be interpreted with care. In particular,
\textit{Basin Tap} appears relatively compact in the t-SNE projection despite
its low recall in the classifier. A compact two-dimensional cluster therefore
does not imply reliable class separation in the original embedding space.
For this reason, we use t-SNE only as qualitative support and base the
quantitative analysis on the confusion matrices and correlation measures in
Section~\ref{sec:bridge}.

The two configurations obtained from the multi-objective search show broadly
similar layouts in Fig.~\ref{fig:tsne_selected}. Both retain the overlap among
the sustained water classes while maintaining more compact regions for the
impulsive events. This similarity indicates that the main embedding pattern
is reasonably stable across the two selected operating points.

\subsection{Ablation Study}
Table~\ref{tab:ablation} examines the contribution of the main architectural
and training choices.

\paragraph{A1: Front-end design.}
Replacing the learned sinc front end with a fixed sinc bank reduces macro-F1
from 0.651 to 0.644, while a matched trainable convolution gives 0.615. The
MFCC front end reaches 0.636. The largest reduction in this group therefore
occurs when the band-pass constraint is removed. The differences should,
however, be interpreted in relation to the variation across seeds; they
support the use of the structured front end but do not imply that every
pairwise difference is statistically significant.

\paragraph{A2: Number of sinc filters.}
Increasing $N_f$ improves mean macro-F1 from 0.613 at $N_f=16$ to 0.628 at
$N_f=32$ and 0.663 at $N_f=64$. Increasing the filter count further to 128
gives 0.672. The improvement therefore becomes smaller beyond 64 filters,
while the front-end width and computational cost continue to increase.

\paragraph{A3: Convolutional body.}
Replacing the depthwise-separable body with standard convolutions gives a
macro-F1 of 0.658 compared with 0.651 for A0. This small difference is well
within the observed seed-to-seed variation, whereas the standard
convolutions require substantially more parameters. The separable body is
therefore retained for the compact models.

\paragraph{A4 and A5: Kernel length and augmentation.}
Macro-F1 increases from 0.638 at $L=101$ to 0.652 at $L=251$ and 0.657 at
$L=401$, indicating a modest benefit from the longer sinc kernel. Removing
amplitude augmentation reduces macro-F1 to 0.625. This result is consistent
with the variation in signal level introduced by different recording
distances and environments.

Overall, the ablation results place most of the sensitivity in the front-end
configuration and amplitude augmentation. Changes to the convolutional body
have a comparatively small effect. The best mean macro-F1 in the ablation
table is 0.672 at $N_f=128$, but the improvement over A0 is modest relative
to the additional filter count.

\begin{table}[H]
	\centering
	\footnotesize
	\setlength{\tabcolsep}{3.5pt}
	\caption{Component ablations under environment-disjoint splitting
		(macro-F1, mean\,$\pm$\,standard deviation over three seeds).
		$\Delta$ denotes the change in mean macro-F1 relative to A0.
		Best macro-F1 is shown in \textbf{bold}.}
	\label{tab:ablation}
	\renewcommand{\arraystretch}{1.12}
	
	\begin{tabular}{llcc}
		\toprule
		\textbf{ID} & \textbf{Variant} &
		\textbf{Macro-F1} & \textbf{$\Delta$ vs. A0} \\
		\midrule
		
		A0  & SincDPNet (reference: $N_f=25$, $L=401$)
		& 0.651\,$\pm$\,0.049 & 0.000 \\
		
		A1a & Fixed (non-learnable) sinc front-end
		& 0.644\,$\pm$\,0.046 & $-0.007$ \\
		
		A1b & Plain trainable conv front-end
		& 0.615\,$\pm$\,0.051 & $-0.036$ \\
		
		A1c & MFCC + same body
		& 0.636\,$\pm$\,0.044 & $-0.015$ \\
		
		\midrule
		
		A2a & $N_f=16$
		& 0.613\,$\pm$\,0.052 & $-0.038$ \\
		
		A2b & $N_f=32$
		& 0.628\,$\pm$\,0.047 & $-0.023$ \\
		
		A2c & $N_f=64$
		& 0.663\,$\pm$\,0.049 & $+0.012$ \\
		
		A2d & $N_f=128$
		& \textbf{0.672\,$\pm$\,0.046} & $+0.021$ \\
		
		\midrule
		
		A3  & Standard conv body (no DS-conv)
		& 0.658\,$\pm$\,0.048 & $+0.007$ \\
		
		A4a & Kernel length $L=101$
		& 0.638\,$\pm$\,0.050 & $-0.013$ \\
		
		A4b & Kernel length $L=251$
		& 0.652\,$\pm$\,0.047 & $+0.001$ \\
		
		A4c & Kernel length $L=401$
		& 0.657\,$\pm$\,0.049 & $+0.006$ \\
		
		A5  & Without amplitude augmentation
		& 0.625\,$\pm$\,0.054 & $-0.026$ \\
		
		\bottomrule
	\end{tabular}
\end{table}
\FloatBarrier

\subsection{Learned Filter Interpretation}
\label{sec:filters}
\begin{figure}[!htbp]
    \centering
    \includegraphics[width=0.35\linewidth]{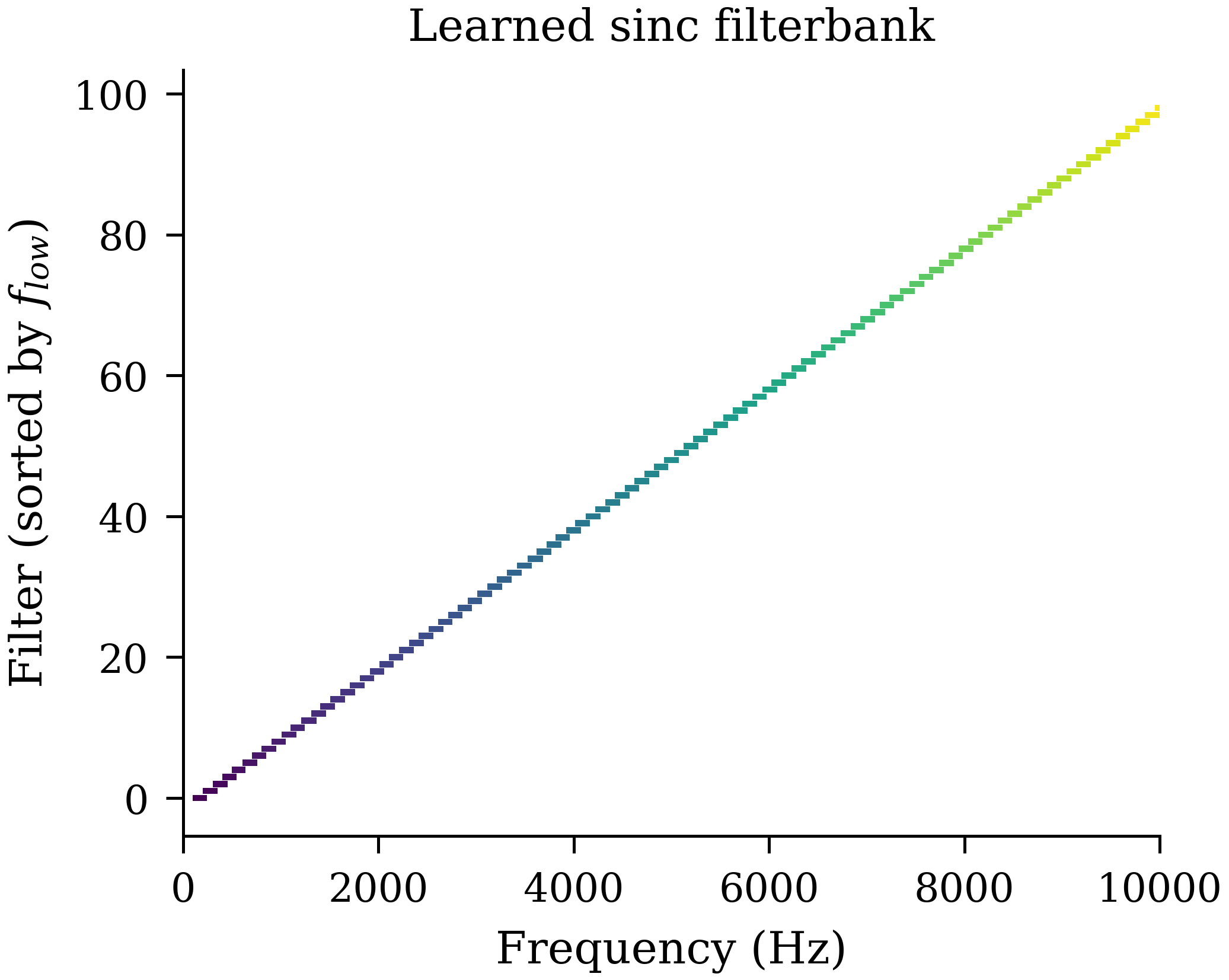}
    \caption{Learned SincDPNet filter bank (each bar is one band-pass filter;
        width $=$ bandwidth). The learned allocation of spectral resolution is
        directly readable in Hz.}
    \label{fig:filterbank}
\end{figure}

\begin{figure}[!htbp]
    \centering
    \includegraphics[width=0.75\linewidth]{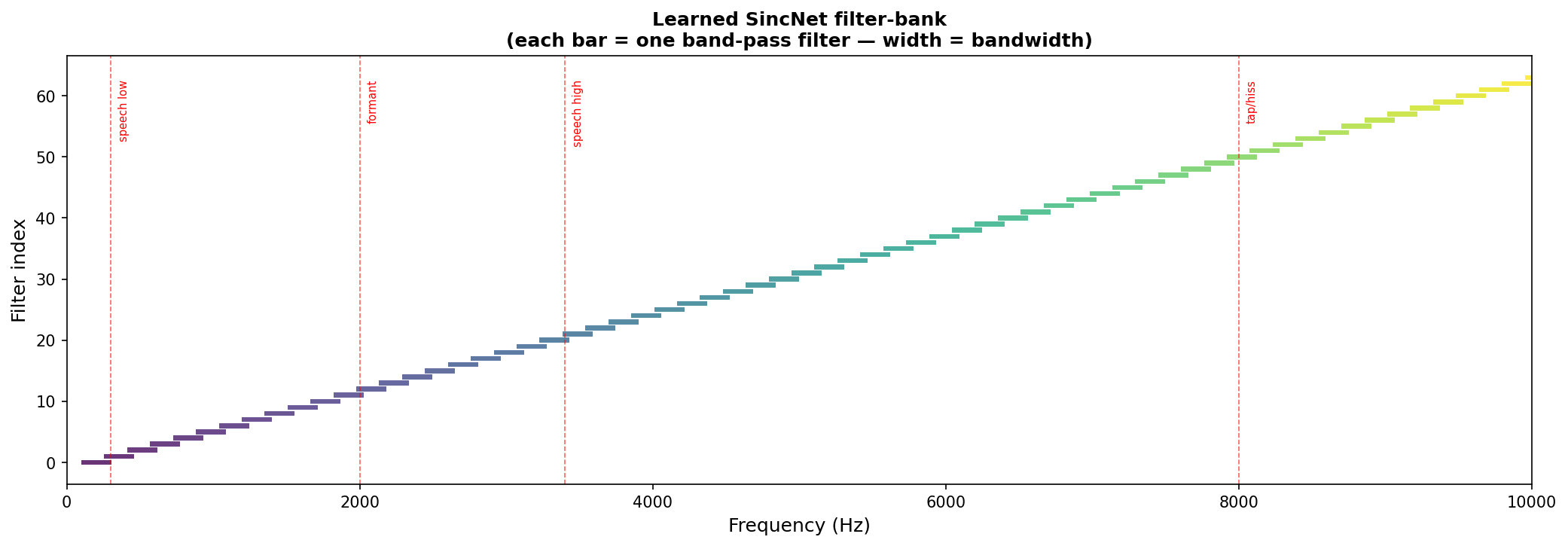}
    \caption{Overlaid magnitude responses of the full learned filter bank on
        a shared frequency axis. The dense overlap in the low-to-mid band and
        the sparse coverage of the high band show, in a single view, how the
        bank concentrates spectral resolution where the water classes carry
        their energy.}
    \label{fig:sincbank}
\end{figure}

\begin{figure}[!htbp]
    \centering
    \includegraphics[width=\linewidth]{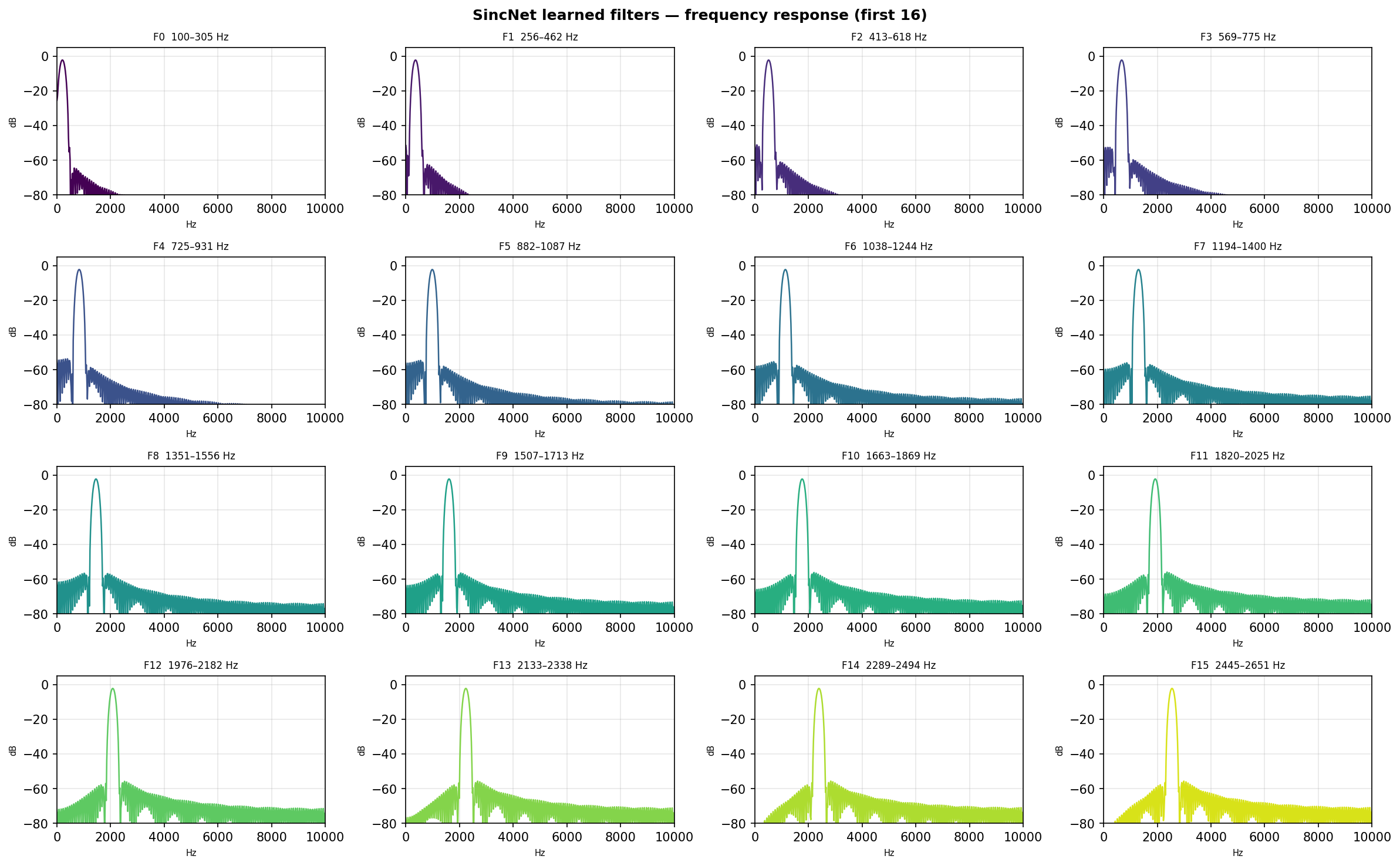}
    \caption{Frequency-magnitude responses of the learned sinc filters. The
        band-pass channels tile the low-to-mid frequency range densely and the
        high range sparsely, the frequency-domain view of the allocation in
        Fig.~\ref{fig:filterbank}.}
    \label{fig:sincfreq}
\end{figure}

\begin{figure}[!htbp]
    \centering
    \includegraphics[width=0.52\linewidth]{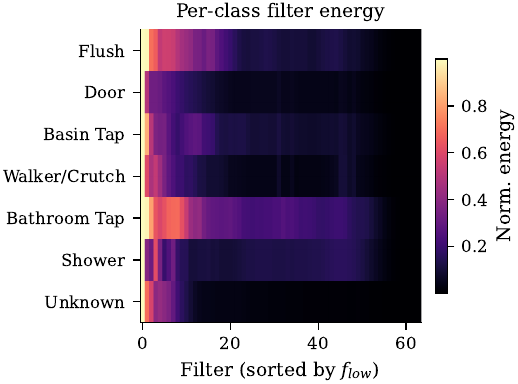}
    \caption{Per-class mean activation energy across learned filters.
        Sustained water classes share active bands (explaining their
        confusability), whereas \textit{Basin Tap} and impulsive classes occupy
        distinct bands.}
    \label{fig:filterenergy}
\end{figure}

\begin{figure}[!htbp]
    \centering
    \includegraphics[width=0.92\linewidth]{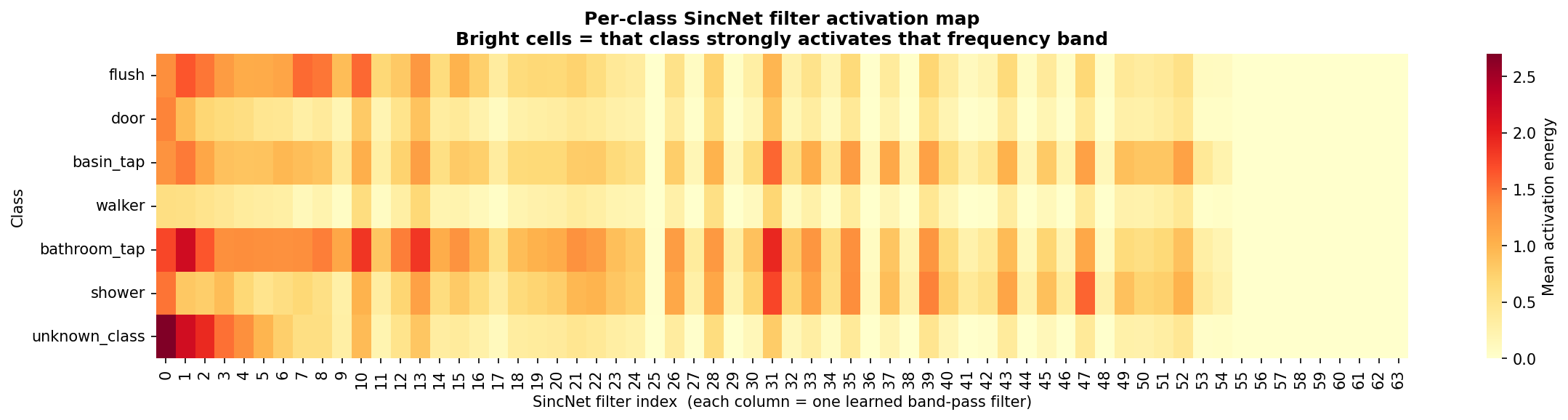}
    \caption{Per-class activation traced across individual learned filters.
        The sustained water classes activate nearly the same filters, while the
        impulsive classes and \textit{Basin Tap} each drive a separable group,
        mirroring the confusion structure of Fig.~\ref{fig:confusion_selected}.}
    \label{fig:sincact}
\end{figure}

The ablation results establish the contribution of the front end; we now inspect
what the learned sinc bank represents. Figures~\ref{fig:filterbank},~\ref{fig:sincbank},
and~\ref{fig:sincfreq} show the learned passbands and their frequency responses, while
Figs.~\ref{fig:filterenergy} and~\ref{fig:sincact} show how the filters respond
across classes.

These filter-level observations motivate the next analysis, which tests whether
the acoustic overlap identified before training is reflected quantitatively in
the model errors.

\FloatBarrier
\subsection{The Interpretability Bridge}
\label{sec:bridge}
We next examine whether the acoustic structure observed before training is
reflected in the errors of the trained model. Two comparisons are used.

At the class level, the difficulty score $D_c$ has little association with
the error rate of the compact $N_f=25$ model
(Spearman $\rho=-0.07$ over seven classes). The scalar score is therefore
not a reliable predictor of class-wise performance. At the pairwise level,
the Bhattacharyya overlap has a modest positive association with
off-diagonal confusion ($\rho=0.39$). The larger confusions occur mainly
among acoustically similar water-flow classes and between
\textit{Door} and \textit{Walker/Crutch}.

\begin{figure}[!htbp]
    \centering
    \includegraphics[width=\linewidth]{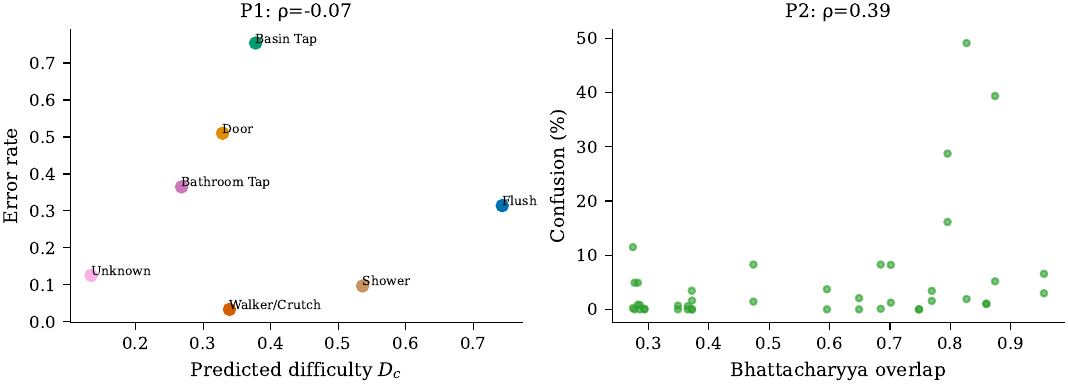}
    \caption{Acoustic-to-model comparison for the compact $N_f=25$ model. Left: the aggregate class-difficulty score has little association with class error. Right: pairwise acoustic overlap has a modest positive association with off-diagonal confusion.}
    \label{fig:bridge}
\end{figure}

Figure~\ref{fig:bridge} summarizes these results. The correspondence between
acoustic overlap, learned-filter responses, and confusion is useful at the
pairwise level, although it does not account for every classification error.
Temporal structure remains important for transient events.

We also vary the five weights used to construct the difficulty score. The
class ordering changes under these perturbations and no stable positive
class-level correlation is obtained. We therefore retain the scalar score
as a descriptive summary of the acoustic analysis and rely on pairwise
overlap when relating the signal statistics to model confusions.

\subsection{Confusion Structure and Error Characterization}
\label{sec:errorchar}
\begin{figure}[!htbp]
    \centering
    \begin{subfigure}[t]{0.49\textwidth}
        \centering
        \includegraphics[width=\linewidth]{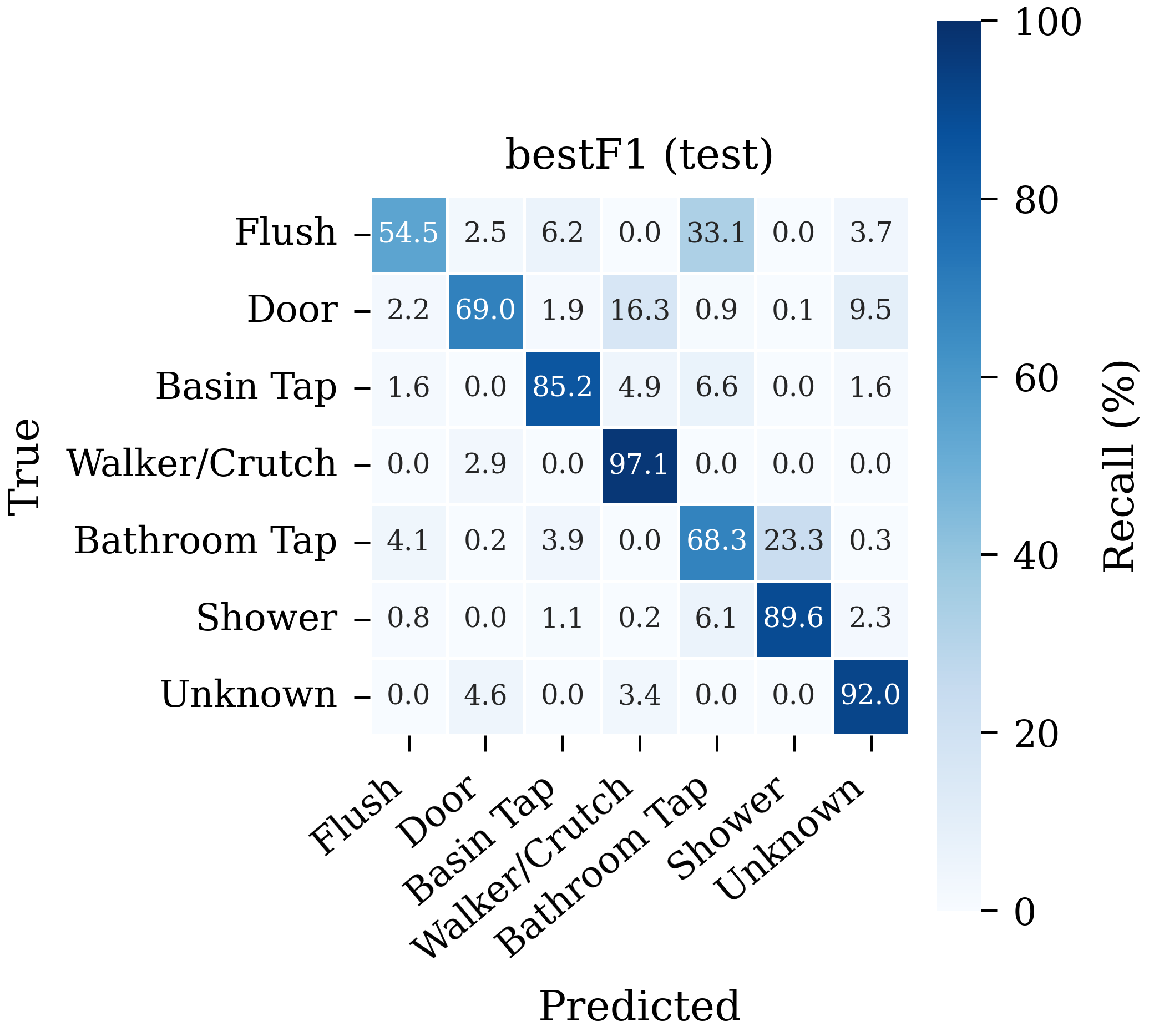}
        \caption{Best-macro-F1 model ($N_f=59$, $\alpha=10$, $5$ blocks; 54.8~KB).}
        \label{fig:conf_bestf1}
    \end{subfigure}\hfill
    \begin{subfigure}[t]{0.49\textwidth}
        \centering
        \includegraphics[width=\linewidth]{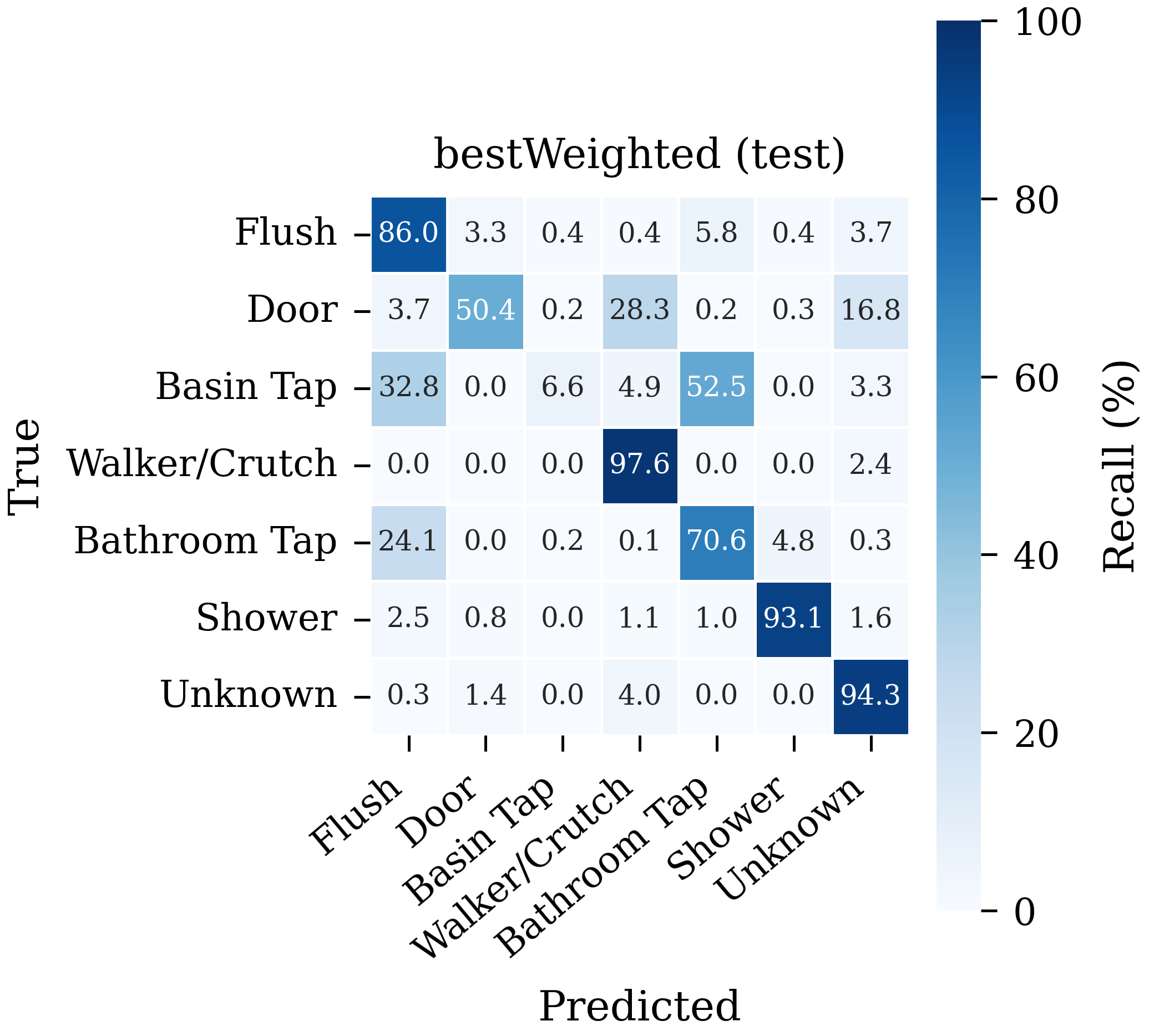}
        \caption{Rank-1 weighted-score model ($N_f=78$, $\alpha=4$, $5$ blocks; 13.3~KB).}
        \label{fig:conf_bestw}
    \end{subfigure}
    \caption{Row-normalized (recall, \%) confusion matrices of the two
        configurations selected by the multi-objective search. The best-F1
        model (a) recovers \textit{Basin Tap} and \textit{Door}, the classes the
        compact model sacrifices, at the cost of $12\times$ more
        footprint; the rank-1 weighted model (b) preserves the easy classes at
        13.3~KB but leaves \textit{Basin Tap} unresolved. Added capacity is
        spent precisely on the hardest, most overlapping classes.}
    \label{fig:confusion_selected}
\end{figure}

We examine the class-wise errors of the compact $N_f=25$ SincDPNet on the
held-out test environment. Figure~\ref{fig:confusion_selected} also compares
the two configurations selected by the multi-objective search. The resulting
error pattern is partly consistent with the spectral analysis, while several
errors reveal the importance of temporal structure.

\begin{figure}[!htbp]
    \centering
    \includegraphics[width=0.6\linewidth]{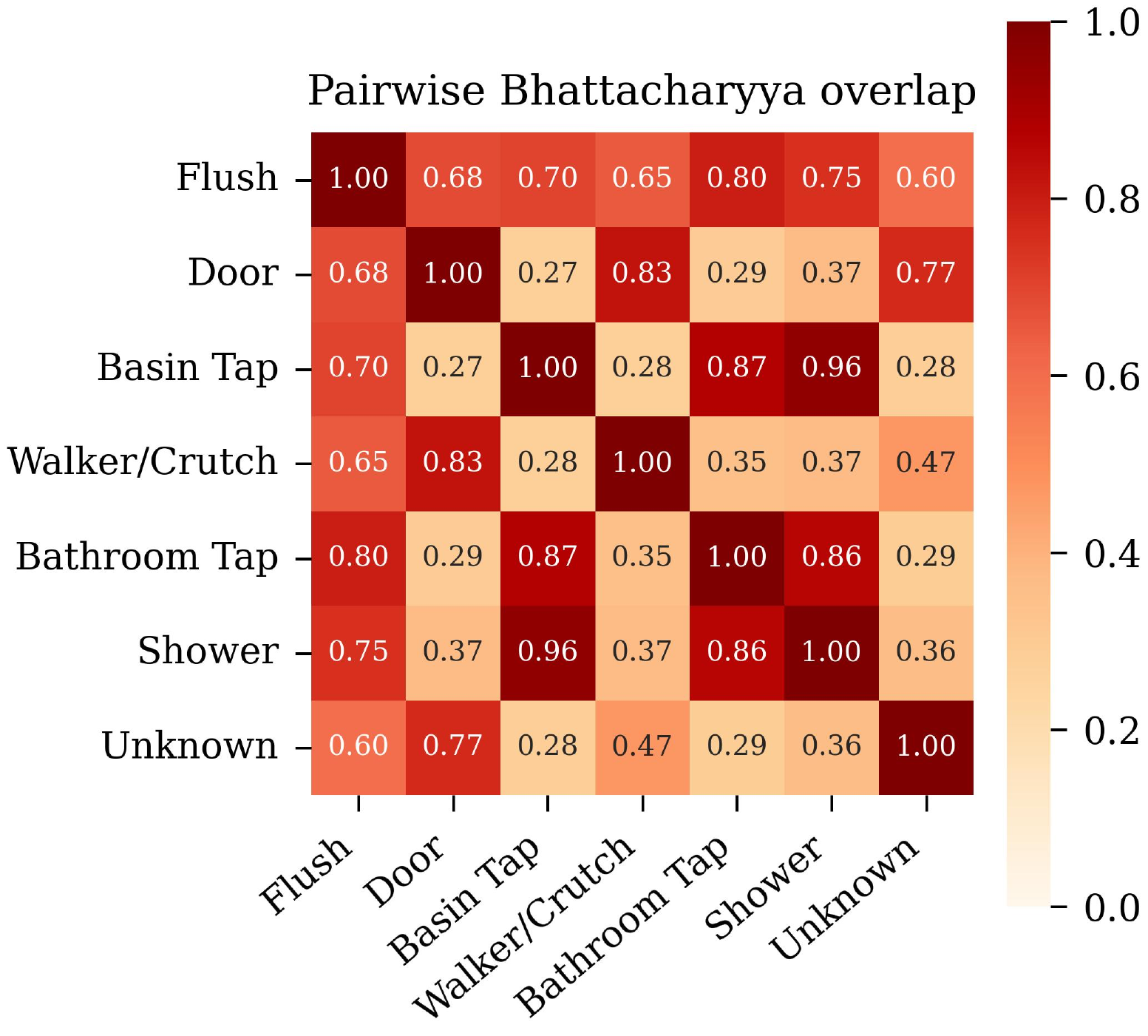}
    \caption{Pairwise Bhattacharyya overlap (Eq.~\ref{eq:bc}) between class
        feature distributions ($1=$ identical). The sustained water classes form
        the highest-overlap block, providing the \emph{a priori} prediction that
        the trained model later confirms in its confusion matrix
        (Fig.~\ref{fig:confusion_selected}).}
    \label{fig:bhattacharyya}
\end{figure}

The compact model achieves high recall for \textit{Walker/Crutch} (97.6\%),
\textit{Shower} (89.4\%), and \textit{Unknown} (88.0\%), with
\textit{Bathroom Tap} reaching 77.0\%. The lowest recalls occur for
\textit{Basin Tap} (8.2\%) and \textit{Door} (24.9\%). Most
\textit{Basin Tap} errors are assigned to \textit{Bathroom Tap} (55.7\%) or
\textit{Shower} (23.0\%), consistent with the overlap among water-related
sounds. \textit{Flush} reaches 61.6\% recall and is most often confused with
\textit{Bathroom Tap} (23.6\%). Figure~\ref{fig:bhattacharyya} provides the corresponding pairwise overlap matrix.

A different mechanism appears for \textit{Door}. It is classified as
\textit{Walker/Crutch} in 67.7\% of cases, whereas only 1.0\% of
\textit{Walker/Crutch} samples are classified as \textit{Door}. Both classes
contain short broadband impacts, but walker or crutch recordings often
contain repeated impacts. A single impact from such a sequence can resemble
a door event. The strong asymmetry therefore cannot be accounted for by
average spectral overlap alone and points to temporal organization and event
boundaries as additional sources of information.

The higher-capacity best-F1 model substantially improves the two weakest
classes of the compact model, raising \textit{Basin Tap} recall from 8.2\%
to 85.2\% and \textit{Door} recall from 24.9\% to 69.0\%. The smaller
rank-1 configuration retains high recall for
\textit{Walker/Crutch}, \textit{Unknown}, \textit{Shower}, and
\textit{Flush}, but remains weak on \textit{Basin Tap}. These results show
where the additional capacity is used: most of the gain occurs for classes
that are poorly separated by the compact representation.

\subsection{Comparison Against the Strongest Baseline}
Among the baseline models, SincNet obtains the highest macro-F1
(Table~\ref{tab:benchmark}), making it a useful comparison for the compact
SincDPNet. SincNet uses approximately $28\times$ more parameters and obtains
higher recall for \textit{Flush} (88.8\% versus 61.6\%),
\textit{Door} (73.6\% versus 24.9\%), and
\textit{Basin Tap} (83.6\% versus 8.2\%). The compact SincDPNet gives
slightly higher recall for \textit{Bathroom Tap} (77.0\% versus 75.6\%) and
\textit{Shower} (89.4\% versus 78.0\%), while the two models perform
similarly on \textit{Unknown}. Both recognize \textit{Walker/Crutch}
reliably.

Despite the difference in model size, the main confusion regions are similar.
Water-related classes remain difficult to separate, and
\textit{Door}/\textit{Walker/Crutch} confusion is present in both models.
The larger SincNet reduces several of these errors, but does not eliminate
the underlying pattern.

\begin{figure}[!htbp]
    \centering
    \includegraphics[width=\linewidth]{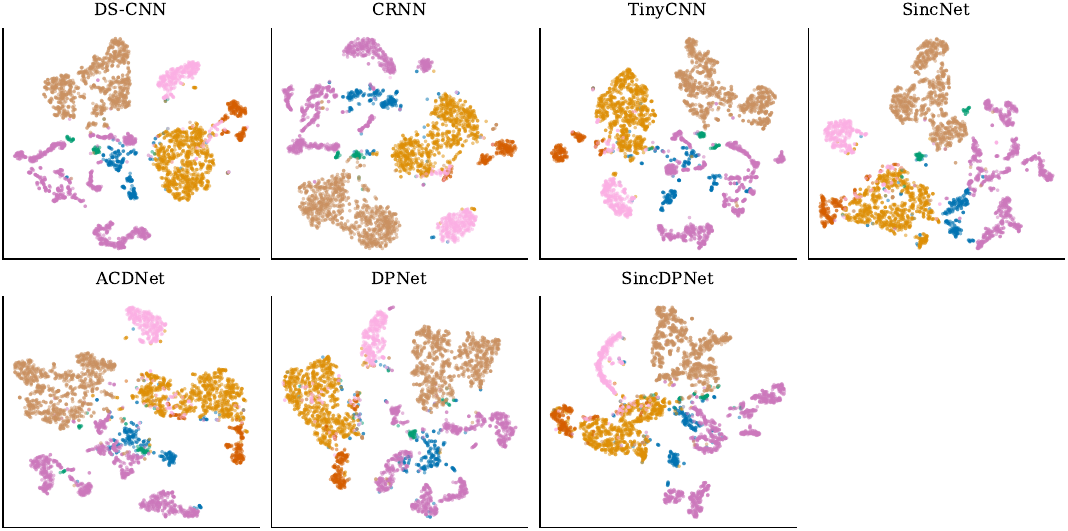}
    \caption{t-SNE embeddings of the held-out test set for all benchmarked
        models. The sustained water classes form a single overlapping region in
        every model regardless of size, while the impulsive classes and
        \textit{Basin Tap} stay separated, evidence that the confusability
        structure is a property of the data, not of any one architecture.}
    \label{fig:tsneall}
\end{figure}

Figure~\ref{fig:tsneall} illustrates the comparative visualization of the embedding
spaces. It can be noted that partial overlapping of sustained water sounds
occurs in all architectures, whereas transient categories tend to have
smaller clusters. Such regularity across all tested architectures is in line
with the acoustic characteristics of the dataset, although this
two-dimensional mapping does not prove the cause of this effect.

Together, the benchmark, embedding, and error analyses identify the operating
points that merit hardware evaluation. The next section therefore turns from
recognition behaviour to deployment-aware model selection.

\FloatBarrier
\section{Deployment-Aware Selection and Hardware Evaluation}
\label{sec:deployment}

The architecture search in Study~A produces a set of solutions with different
trade-offs between recognition performance and model footprint. Selecting only
the configuration with the highest validation macro-F1 may result in unnecessary
memory and computational overhead, whereas selecting only the smallest model may
lead to an unacceptable reduction in recognition performance. We therefore
perform an additional deployment-aware analysis using three representative
configurations: (i) the highest weighted-score model, which balances validation
macro-F1 and model size; (ii) the highest-validation-F1 model, representing the
accuracy-oriented operating point; and (iii) a compact Pareto-optimal model,
representing the deployment-oriented operating point.

The weighted ranking in Study~A is obtained after min--max normalization of
validation macro-F1 and model-size efficiency. The normalized objectives are
combined as
\begin{equation}
    S_i =
    0.7
    \frac{F_i-F_{\min}}{F_{\max}-F_{\min}}
    +
    0.3
    \left(
    1-
    \frac{M_i-M_{\min}}{M_{\max}-M_{\min}}
    \right),
    \label{eq:deployment_weighted_score}
\end{equation}
where $F_i$ and $M_i$ denote the validation macro-F1 and model size of the
$i$th configuration, respectively. The first term favours recognition quality,
whereas the second term assigns a higher score to smaller models. Configurations
are ranked in descending order of $S_i$.

Table~\ref{tab:deployment_candidates} summarizes the three configurations
selected for detailed deployment analysis. The highest weighted-score model uses
$N_f=78$ Sinc filters, a width multiplier of $\alpha=4$, and five
depthwise-separable blocks. It achieves a  validation macro-F1 (search-time) of
$0.8443$ with a recorded search-time model size of $13.3$~KB. The
accuracy-oriented model uses $N_f=59$, $\alpha=10$, and five blocks, and obtains
the highest validation macro-F1 of $0.8942$, although its recorded
size increases to $54.8$~KB.

The compact Pareto configuration uses only $N_f=25$ filters, $\alpha=6$, and
four blocks. Its validation macro-F1 is $0.8016$, which is lower than that of
the accuracy-oriented model but is obtained with a substantially smaller
recorded footprint of $11.1$~KB. This configuration is retained because no
other evaluated solution simultaneously provides a higher macro-F1 and a
smaller search-time model size.

\begin{table}[t]
    \centering
    \caption{Study-A configurations selected for deployment-aware evaluation.
        The three configurations represent distinct points on the Pareto frontier:
        best weighted score (rank-1), highest validation macro-F1, and a compact model.}
    \label{tab:deployment_candidates}
    
    \small
    \setlength{\tabcolsep}{5.5pt}
    \renewcommand{\arraystretch}{1.15}
    
    \begin{tabular}{@{}lcccccc@{}}
        \toprule
        \textbf{Selection role} &
        \textbf{$N_f$} &
        \textbf{$\alpha$} &
        \textbf{Blocks} &
        \textbf{Val.\ macro-F1} &
        \textbf{Size (KB)} \\
        \midrule
        
        Highest weighted score (rank-1) &
        78  & 4  & 5 & 0.8443 & 13.3 \\
        
        Highest validation F1 &
        59  & 10 & 5 & \textbf{0.8942} & 54.8 \\
        
        Compact Pareto model &
        25  & 6  & 4 & 0.8016 & 11.1 \\
        
        \bottomrule
    \end{tabular}
    
    \vspace{3pt}
    
\end{table}

Before timing these configurations on hardware, we first verify that deployment
conversion preserves their predictions.

\subsection{Freezing and TFLite Conversion}

The learned Sinc filters are first converted into an equivalent fixed
convolutional kernel before TensorFlow Lite conversion. This removes the
analytical filter-construction operations from the inference graph while
preserving the learned frequency responses. Prediction agreement between the
original Keras model and the frozen model is $100\%$ for all three completed
configurations. The frozen Keras and FP32 TensorFlow Lite models also reproduce
the original Keras predictions exactly on the complete held-out test set.

This result is important because it separates errors caused by graph conversion
from those introduced by numerical quantization. In the present experiments,
the conversion from the trainable Sinc formulation to a frozen convolution and
then to FP32 TensorFlow Lite introduces no measurable degradation. Therefore,
the FP32 TensorFlow Lite models provide reliable software representations for
subsequent hardware evaluation.

\subsection{Deployment on Embedded Hardware}
\label{sec:rpi-deploy}

In order to evaluate the practical applicability of the models beyond
training, we ran the models on a low-power single board computer and
calculated the inference latency. It should be noted that in this section,
the objective is not to optimize the model performance for a specific
hardware platform, but rather to demonstrate that the proposed interpretable
raw wave form design is still lightweight enough to work on hardware
commonly used in always-on acoustic sensing applications where the
constraints on power and cost are critical. Inference times for one $1.5$~s
audio clip are provided below.

The primary target is the Raspberry Pi Zero~2~W, a low-cost board with a
quad-core $1$~GHz processor and $512$~MB of RAM and no dedicated accelerator,
so all computation runs on the CPU. As a reference point we also time the
compact model on a laptop with an Intel Core~U5 processor. Both run the
identical exported model and the same pre-processing front end, so the only
variable is the hardware. Inference time is measured as the compute time for a
single clip and excludes the fixed $1.5$~s needed to capture the audio, since
capture duration is a property of the task rather than of the model or device.

We use two metrics to describe on-device behaviour. \emph{Inference time} is
the median wall-clock time to classify one $1.5$~s clip, measured on the device
and excluding audio capture. The \emph{real-time factor} (RTF) is the ratio of
this inference time to the clip duration,
\begin{equation}
    \mathrm{RTF} = \frac{t_{\mathrm{inference}}}{t_{\mathrm{clip}}},
    \qquad t_{\mathrm{clip}} = 1.5~\text{s},
    \label{eq:rtf}
\end{equation}
so that $\mathrm{RTF}<1$ means the model classifies a clip faster than the clip
takes to record, leaving the processor idle for part of each cycle, whereas
$\mathrm{RTF}\ge 1$ means inference cannot keep pace with a continuous audio
stream. A low RTF is therefore desirable for always-on monitoring, since the
remaining time budget can be used for audio capture, buffering, and any
post-processing within the same cycle. Table~\ref{tab:rpi-latency} reports the
measured latency and RTF for the baseline and the three selected SincDPNet
configurations on both hardware platforms.


\begin{table}[!t]
    \centering
    \small
    \caption{Single-clip inference performance of DPNet and selected SincDPNet configurations on embedded and reference hardware for a $1.5$~s audio clip. Inference time excludes audio capture; RTF $<1$ indicates faster-than-real-time operation.}
    \label{tab:rpi-latency}
    \setlength{\tabcolsep}{5pt}
	\begin{tabular}{@{}llccccc@{}}
		\toprule
		\textbf{Device (RAM, clock)} & \textbf{Model} & \textbf{Params} &
		\textbf{Acc.\,(\%)} & \textbf{Macro-F1} &
		\textbf{Inf.\ time} & \textbf{RTF} \\
		\midrule
		
		\multirow{4}{*}{\begin{tabular}[c]{@{}l@{}}
				Raspberry Pi Zero~2~W\\
				($512$~MB, $1$~GHz)
		\end{tabular}}
		& DPNet (baseline)
		& 4{,}944
		& 69.2
		& 0.612
		& \textbf{15~ms}
		& \textbf{0.01} \\
		
		& SincDPNet, compact ($N_f{=}25$)
		& \textbf{2{,}848}
		& 75.7
		& 0.661
		& 435.6~ms
		& 0.29 \\
		
		& SincDPNet, best-F1
		& 14{,}040
		& \textbf{80.2}
		& \textbf{0.760}
		& 920.4~ms
		& 0.61 \\
		
		& SincDPNet, rank-1 (weighted)
		& 3{,}408
		& 75.6
		& 0.673
		& 886.3~ms
		& 0.59 \\
		
		\cmidrule(l){1-7}
		
		\multirow{4}{*}{\begin{tabular}[c]{@{}l@{}}
				Laptop (Intel Core~U5)\\
				($16$~GB, $1.2$~GHz)
		\end{tabular}}
		& DPNet (baseline)
		& 4{,}944
		& 69.2
		& 0.612
		& \textbf{2.5~ms}
		& \textbf{0.0016} \\
		
		& SincDPNet, compact ($N_f{=}25$)
		& \textbf{2{,}848}
		& 75.7
		& 0.661
		& 28.0~ms
		& 0.0186 \\
		
		& SincDPNet, best-F1
		& 14{,}040
		& \textbf{80.2}
		& \textbf{0.760}
		& 53.4~ms
		& 0.0356 \\
		
		& SincDPNet, rank-1 (weighted)
		& 3{,}408
		& 75.6
		& 0.673
		& 59.4~ms
		& 0.0396 \\
		
		\bottomrule
	\end{tabular}
\end{table}

All four models achieve a real-time factor well below one on the Raspberry Pi
Zero~2~W, confirming that even the accuracy-oriented configuration is
deployable on this class of hardware, while the compact model leaves the
largest idle margin. This is the deciding information when selecting a model
for embedded acoustic monitoring, where a small loss in accuracy may be
acceptable in exchange for a substantially lower on-device latency.

These hardware results complete the empirical evaluation and provide the basis
for the broader interpretation that follows.

\section{Discussion}
\label{sec:discussion}
There is one common thread that runs through all of the above contributions. On this task, the hard problems come from the data, not the model. The superior model is the one whose architecture matches the problem, rather than its capacity. The signal analysis identifies the hard examples prior to training. The front-end then provides the model with filters tuned to the same frequencies as determined by the analysis. The architectural search and ablation study prove that going beyond a few thousand parameters in capacity provides no benefit. The error analysis reveals the residual errors come from places where the acoustics are inherently ambiguous. The big black box may have equal or superior performance, but it cannot explain itself as this work can. On the privacy-sensitive edge device, the explanation is part of the product. The rest of this section covers each of the above contributions in sequence, but at each point we also make clear what the data do not show.

\dataset{} is unique by virtue of what it highlights rather than by its volume Table~\ref{tab:dataset_comparison} shows the gap it fills. Existing bathroom corpora capture water events, but none also adds door and mobility-aid sounds together with an explicit out-of-distribution class. This is evidenced by the outcome of the study. These impulsive classes are only available in this corpus, and they yield the largest number of errors. As shown in Table~\ref{tab:errors}, this number equals $774$ of $1{,}209$. Without the events of water, this corpus would never reveal this flaw.

\begin{table}[!htbp]
    \centering
    \small
    \setlength{\tabcolsep}{4pt}
    \caption{Error characterization of the deployed SincDPNet over its
        $T=1{,}209$ misclassified test clips, assigned to mutually exclusive
        categories by an automated rule-based audit (each clip is counted once,
        under the first matching rule). Two acoustic categories account for
        $97.6\%$ of all errors.}
    \label{tab:errors}
    \renewcommand{\arraystretch}{1.15}
    \begin{tabular}{lc}
        \toprule
        \textbf{Error category} & \textbf{Count} \\
        \midrule
        Transient/impulsive overlap                     & 774 \\
        Broadband water-flow ambiguity                  & 406 \\
        Non-target/unknown confusion           & 29 \\
        Low SNR / background masking                    & 0 \\
        Annotation-boundary (partial event capture)           & 0 \\
        \midrule
        \textbf{Total}                                  & 1{,}209 \\
        \bottomrule
    \end{tabular}
\end{table}

This dataset is consistent with the analysis performed in this paper. The figures \ref{fig:sigstats} and \ref{fig:temporal} lay the ground for distinguishing the transient vs sustained case using only statistics. The figure~\ref{fig:spectra} depicts the water classes overlap at the middle frequency range. The figure~\ref{fig:bhattacharyya} computes this overlap class by class. The figure~\ref{fig:lda} demonstrates that this overlap persists after the optimal linear projection. Four different ways of looking at this problem give the same result before training a single network.

The benchmark in Table \ref{tab:benchmark} is strong evidence for something beyond ``our model is small.'' Interpreting the results as a function of front-end type rather than rank produces a clear trend: the two raw-waveform models with unconstrained front-ends come up at the bottom (DPNet 0.612 macro-F1, ACDNet 0.639 macro-F1). Meanwhile, the two models with constrained band-pass front-ends appear near the top (SincNet 0.712 macro-F1, SincDPNet 0.661 macro-F1). The input domain does not split the field; rather, it is the front-end structure that does. The most direct contrast in terms of performance comes from SincDPNet vs.\ DPNet: at nearly identical size (2,848 versus 4,944 parameters), the sinc front-end provides a boost of +0.096 macro-F1 and +0.085 MCC. As both models are of the same family, this contrast isolates the role of the constrained front-end.

Equation \eqref{eq:bandpass} explains why this is possible. Each filter has two parameters, not $L$, thus making the whole filterbank equivalent to $2N_f$ parameters, i.e., 128 parameters out of 2,848 (2.8\%) of the model's total parameters (Table \ref{tab:budget}). Equation \eqref{eq:dsratio} explains the cost-effectiveness of the front-end: separable convolutions are $C_{\text{out}} + k^2$ times less expensive than a regular convolution. Taken together, the properties make the model 10-fold cheaper than the 128 k budget of DCASE \citep{martinmorato2022low}, while delivering the second-best MCC result in Table \ref{tab:benchmark}.

The ablation experiment in Table~\ref{tab:ablation} provides support for this observation from within the model instead of across models and further refines it. By disentangling the two effects of the front-end, we see that the band-pass design effect adds 0.014 macro-F1 compared to a regular convolution, and learning the band-pass edges effect adds 0.027 macro-F1 each of which is not sufficient to explain the improvement. Replacing the separable body with standard convolutions adds or subtracts 0.002 in accuracy, which is less than seed noise but adds nine times more parameters. An architectural search adds a different line of evidence: according to Table~\ref{tab:ranking}, the best architecture is 5.1 KB while the biggest one that was explored (65 KB) achieves 0.550 macro-F1. Thus, three independent pieces of evidence—a cross-model benchmark, an within-model ablation, and an architecture search all converge on the conclusion that the binding constraint here is not capacity.
.

This discussion should not be misinterpreted. SincDPNet does not deliver the highest macro-F1. This honor belongs to SincNet, but at 28$\times$ more parameters. The point being made concerns Pareto-efficiency, not dominance. The correct interpretation of Table \ref{tab:benchmark} is that structuring the front-end brings considerable gains when scaled to small size, but it cannot entirely substitute the lack of capacity.

A central result is Figure~\ref{fig:bridge}. We compute a difficulty
score from signal statistics alone, before any training, and compare it
with the trained model's per-class errors. The hardest class in
Table~\ref{tab:difficulty} is \textit{Flush}, and it is indeed among
the worst served ($61.6\%$ recall). The clear exception is \textit{Basin
    Tap}, which the score ranks mid-difficulty yet the model serves worst of
all. We discuss that case below. The overlap-versus-confusion
correlation is $\rho=0.39$. Read together, the two results are
informative. Acoustic overlap explains much of how hard a class is, but
not all of it, and which confusion happens also depends on the model's
temporal resolution.

Table~\ref{tab:errors} makes this concrete, and it corrects a
prediction. We expected water-flow ambiguity to dominate. It is second,
at $406$ errors. The impulsive
\textit{Door}$\leftrightarrow$\textit{Walker/Crutch} confusion is
first, at $774$. Both are short broadband knocks whose useful
information sits in fine temporal detail. The pooling that makes the
body cheap throws some of that detail away. This is a direct cost of the
efficiency built in Section~\ref{sec:body}, and we had not noticed it
before.  The reason why we describe this fact instead of the predicted one is to establish an error taxonomy.
There are two error-free categories. No errors caused by low signal-to-noise ratio and clipped events were found (Table~\ref{tab:errors}). It means that the corpus is properly segmented and recorded, and the rest of errors come from real acoustic ambiguity.

\paragraph{\textbf{Interpretability of the Learned Filter Bank:}}
The interpretability of SincDPNet comes directly from its first layer. 
Figure~\ref{fig:filterbank} shows the frequency ranges learned by the sinc 
filters, while Fig.~\ref{fig:filterenergy} shows how strongly these filters 
respond to each activity class. Clear class-dependent patterns can be observed. 
Impulsive events and \textit{Basin Tap}, for example, concentrate their energy 
in relatively distinct groups of filters, whereas sustained water-related 
activities produce broader responses across the low- and mid-frequency bands. 
These patterns provide a physical explanation for both successful predictions 
and class confusions, rather than relying only on internal feature activations.

This is particularly useful for an assistive monitoring system, where it is 
important to understand not only the overall recognition accuracy but also 
which events are likely to be confused and why. For example, the strong 
recognition of shower, tap, and mobility-aid sounds is consistent with their 
characteristic filter-response patterns. In contrast, the similar responses 
observed for \textit{Door} and \textit{Walker} help explain their higher mutual 
confusion. This suggests that separating these two activities may require 
additional temporal context or a dedicated secondary classifier. Thus, the 
learned sinc filters provide an interpretable link between the acoustic 
characteristics of an event and the final recognition behaviour of the model. 
For reproducibility and further analysis, the learned filter-bank parameters 
are also released with the implementation.

From a privacy-first assistive monitoring point of view, interpretability is of practical use. Not only it takes into account the average F1 score per class but also answers the following two questions: what events may go undetected, and are we able to justify it? In our case, showers, taps, and sounds of mobility aids are detected with a high confidence level, while detection of the \textit{Door}/\textit{Walker} pair requires either longer time window or special sub-classification. These observations follow from interpretation of particular frequency bands rather than from non-transparent activation patterns. This focus on interpretability is our rationale for publishing learned filter banks.

\paragraph{\textbf{The MOO-based Design Procedure:}}
We claim no methodological novelty in the optimization, so the relevant
question is whether our reported design depends on optimizer settings.
Table~\ref{tab:bo-ablation} answers it. The four surrogate-guided
variants finish within $0.087\%$ of one another in hypervolume. The gap
to random search is $0.61\%$, about seven times larger.
Figure~\ref{fig:hv} shows the same ordering in the trajectories. Every
surrogate variant reaches $90\%$ of its hypervolume gain within $22$
evaluations, and random search does not reach that within the budget. So
a reader who repeats this study with a different acquisition function
should get a similar design. One point is worth stating plainly: on a
space this small, Bayesian optimization buys convergence speed and
reproducibility, not a dramatically better front.


\paragraph{\textbf{Broader Implications:}}
Even though this research addresses the recognition of activities in the
bathroom setting, the presented framework can be applied for other similar
small audio classification problems including recognition of household
activities, monitoring condition of machines, and classification of animal
sounds \cite{Garai2026}. The above mentioned applications share similar
characteristics related to small training datasets and classes that have
spectral and/or temporal structures.

On the whole, the work illustrates an easy-to-use design rule: study the separability of classes prior to training, incorporate the result into the architecture of the front end, and test using splits that capture the principal variance, which could be the room, device, site, subject, or recording session. This will enable you to understand whether your system has captured the desired acoustic events or other characteristics of the recording environment.
These findings provide us with the results outlined below.

\section{Conclusion}
\label{sec:conclusion}

In this paper, \dataset{} and SincDPNet are introduced for bathroom acoustic event classification from waveforms. For the experiment, recordings are divided into sessions and environments before overlapping segments are made, which avoids having correlated segments across training, validation, and testing. Under this environment-disjoint protocol, the compact $N_f=25$ SincDPNet achieves 75.7\% accuracy, 0.661 macro-F1, and 0.716 MCC with 2,848 trainable parameters.

The main contribution is a model that brings together compact raw waveforms and physically interpretable acoustic features. In sinc front-end, there are just $2N_f$ learnable frequency variables (say, $50$ in the compact $N_f=25$ model) and the passbands are immediately readable in hertz. Pairwise acoustic decomposition, as well as the confusion matrix, indicate the same difficult-to-classify groups: similar water flow acoustic samples, and the two impulse classes. Class difficulty measure cannot reliably predict the error order in all cases; however, pairwise class overlap reveals more.

The multi-objective search spans 24 evaluated configurations. The highest weighted-score configuration uses $N_f=78$ and reaches macro-F1 0.8443 at 13.3~KB, while the highest validation macro-F1, 0.8942, is obtained by a larger $N_f=59$ configuration at 54.8~KB. These results show why a Pareto view is useful: the most accurate design and the most balanced design are different. From this Pareto set we analyze three representative operating points, including the compact $N_f=25$ model used for the detailed error and deployment analysis.

Future work will focus on three directions. First, the dominant
\textit{Door}/\textit{Walker} confusion will be addressed by preserving
finer temporal information, either through longer temporal context or a
lightweight secondary classifier for acoustically similar impulsive events.
Second, the generalizability of the proposed framework will be examined across
a broader range of rooms with different acoustic characteristics, recording
devices, and acquisition conditions. This evaluation will assess the extent to
which the learned sinc bands remain transferable beyond the present corpus.
Finally, the selected Pareto-optimal models will be deployed on more
resource-constrained platforms, including microcontroller-class hardware, to
measure memory usage, inference latency, and energy consumption directly under
deployment conditions.
\section*{Data and Code Availability}

The dataset used in this study can be requested through the
\href{https://docs.google.com/forms/d/e/1FAIpQLSeG6SI1RTFGJUxEYZn49DJKLeNmeOQVrni4vGBuXhI87fPw9w/viewform?usp=dialog}
{Dataset Access Form}. The source code, model configurations, and experimental
scripts used in this study are publicly available at
\url{https://github.com/debolina-34/SnaanGhar7}.
\section*{Ethics Statement}
All participants voluntarily consented to the collection and public release of anonymized recordings for academic research; property-owner permission was obtained for each site.

\section*{Acknowledgments}
The authors thank the participating households and institutions for
granting access to their facilities for data collection.

\FloatBarrier
\bibliographystyle{plainnat}
\bibliography{refs_sincdpnet}

\end{document}